\documentclass[12pt,a4paper]{article}
\usepackage[round]{natbib}
\usepackage[affil-it]{authblk}
\newcommand{\q}[1]{``#1''}
\usepackage{mathrsfs}
\usepackage{amsmath}
\usepackage{booktabs}
\usepackage{multirow}
\usepackage{siunitx}
\usepackage{enumerate}
\usepackage{setspace}
\usepackage[utf8]{inputenc}
\usepackage[english]{babel}
\usepackage{amsthm}
\usepackage{amssymb}
\usepackage{extarrows}
\usepackage{changepage}
\usepackage{yhmath}
\usepackage{color}
\usepackage{rotating}
\usepackage{algorithm}
\usepackage[noend]{algpseudocode}
\usepackage[title]{appendix}
\def\BState{\State\hskip-\ALG@thistlm}
\usepackage{enumitem}

\usepackage{graphicx}
\usepackage{float}
\usepackage{caption}
\usepackage{subcaption}
\usepackage{geometry}

\makeatother
\makeatletter
\usepackage[rightcaption]{sidecap}

\graphicspath{ {images/} }

\makeatletter

\newcommand{\Romannum}[1]{\expandafter\@slowromancap\romannumeral #1@}
\makeatother

\providecommand{\keywords}[1]
{
  \small	
  \textbf{\textit{Keywords:}} #1
}

\title{An Ensemble Method for Interval-Censored Time-to-Event Data}
\author[1]{Weichi Yao}
\affil[1]{Stern School of Business, New York University \authorcr
  Email: wy635@stern.nyu.edu}
\author[2]{Halina Frydman}
\affil[2]{Stern School of Business, New York University \authorcr
  Email: hfrydman@stern.nyu.edu} 
\author[3]{Jeffrey S. Simonoff}
\affil[3]{Stern School of Business, New York University \authorcr
  Email: jsimonof@stern.nyu.edu} 
\date{\today}
\begin{document}
\maketitle
\begin{abstract}
Interval-censored data analysis is important in biomedical statistics for any type of time-to-event response where the time of response is not known exactly, but rather only known to occur between two assessment times. Many clinical trials and longitudinal studies generate interval-censored data; one common example occurs in medical studies that entail periodic follow-up. In this paper we propose a survival forest method for interval-censored data based on the conditional inference framework. We describe how this framework can be adapted to the situation of interval-censored data. We show that the tuning parameters have a non-negligible effect on the survival forest performance and guidance is provided on how to tune the parameters in a data-dependent way to improve the overall performance of the method. Using Monte Carlo simulations we find that the proposed survival forest is at least as effective as a survival tree method when the underlying model has a tree structure, performs similarly to an interval-censored Cox proportional hazards model fit when the true relationship is linear, and outperforms the survival tree method and Cox model when the true relationship is nonlinear. We illustrate the application of the method on a tooth emergence data set.\\

\end{abstract}

\section{Introduction}
Most statistical methods for the analysis of survival time (time-to-event) data have been developed in the situation where the observations could be right-censored. In many situations, however, the survival time cannot be directly observed and it is only known to have occurred in an interval obtained from a sequence of examination times. In this situation, we say that the survival time is interval-censored. 

Interval-censored data are encountered in many medical and longitudinal studies and various methods have been developed for their analysis. \citet{Fcox} provided the first method for estimation of the Cox proportional hazard model from interval-censored data. Surveys of later approaches to the estimation of the Cox model and other semi or parametric survival models for interval-censored data can be found in \citet{SunIC} and \citet{new}. However, these methods rely on restrictive assumptions such as proportional hazards and a log-linear relationship between the hazard function and covariates. Furthermore, because these methods are often parametric, nonlinear effects of variables must be modeled by transformations or expanding the design matrix to include specialized basis functions for more complex data structures in real world applications.

Recently, \citet{ICtree} proposed a nonparametric recursive-partitioning (tree) method for interval-censored survival data, as an extension of the conditional inference tree method for right-censored data of \citet{ctree}. As is well known, tree estimators are nonparametric and as such often exhibit low bias and high variance. Compared to simple models like trees, ensemble methods like bagging and random forest can reduce variance while preserving low bias. These methods average over predictions of the base learners (the trees) that have been fit to bootstrap samples, and are able to remain stable in high-dimensional settings and therefore can substantially improve prediction performance \citep{RF}. \citet{RSF} proposed the random survival forest (RSF) that extends random forest \citep{RF} to right-censored survival data. \citet{SE} proposed the conditional inference survival forest (with the conditional inference survival tree as the base learner) by incorporating weights into random forest-like algorithms and extending gradient boosting in order to minimize a weighted form of the empirical risk.

In this paper, we propose a conditional inference survival forest method appropriate for interval-censored data (we will refer to this method as the IC cforest method). The goal of this ensemble tree algorithm is to lower the variance compared to an individual tree and therefore stabilize and improve the prediction performance. The proposed method is an extension of the conditional inference forest method (which is designed to handle right-censored survival data, and will be referred as the cforest method) with the base learner being the conditional inference survival tree proposed by \citet{ICtree} (we will refer to this as the IC ctree method).
\section{An interval-censored survival forest}
\subsection{Extending the survival forest of \citet{SE}}\label{sec:excf}
The recursive partitioning proposed in \citet{ctree} for building the ctree is based on a test of the global null hypothesis of independence between response variable $\mathbf{Y}$ and any of the $m$ covariates $X_1,...,X_m$. As a decision tree-based ensemble method, cforest induces randomness into each node of each individual tree (that is built from a bootstrap sample) when selecting a variable to split on. Only a random subset of covariates is considered for splitting at each node. 
The recursive partitioning in cforest is based on a test of the global null hypothesis of independence between response variable $\mathbf{Y}$ and any of the elements in a random subset $I$ of the total $m$ covariates (indeed, the size of this random subset $|I|$ is prespecified, with further discussion given in Section \ref{sec:OOBdes}). In each node, after such a random subset $I$ is selected, permutation-based multiple testing procedures are applied. The recursion stops if the global null hypothesis of independence cannot be rejected at a prespecified level $\alpha$. If it can be rejected, the association between $\mathbf{Y}$ and each of the covariates $X_j$, $j\in I$ is measured to select the covariate with strongest association to the response variable $\mathbf{Y}$ (the one with minimum $p$-value, indicating the largest deviation from the partial null hypotheses). Once a covariate is selected, the permutation test framework is again used to find the optimal binary split.

The $|I|$-dimensional covariate vector $\mathbf{X}_I = (X_j)_{j\in I}$ falls in a space denoted by $\mathcal{X}_I=\prod_{j\in I}\mathcal{X}_j$, and $\mathbf{Y}\in\mathcal{Y}$. The association of the response variable $\mathbf{Y}$ and a predictor $X_j$, $j\in I$ based on a random sample $\mathcal{L}_n=\{(\mathbf{Y}_i,X_{1i},X_{2i},..., X_{mi});\;i=1,...,n\}$ is measured by linear statistics of the form
\begin{equation*}
T_j(\mathcal{L}_n,\mathbf{w}) = vec\left(\sum_{i=1}^nw_ig_j(X_{ji})h(\mathbf{Y}_i,(\mathbf{Y}_1,...,\mathbf{Y}_n))^T\right)\in\mathbb{R}^{p_jq},
\end{equation*}
where $\mathbf{w}:=(w_1,...,w_n)$ is a vector of non-negative integer-valued case weights having nonzero elements when the corresponding observations are elements of the node and zero otherwise, $g_j:\mathcal{X}_j\rightarrow\mathbb{R}^{p_j}$ is a nonrandom transformation of covariate $X_j$, and $h: \mathcal{Y}\times\mathcal{Y}^n\rightarrow\mathbb{R}^q$ is the influence function and depends on the responses $(\mathbf{Y}_1,...,\mathbf{Y}_n)$ in a permutation-symmetric way. The dimension $p_j$, $j\in I$ and $q$ vary according to different practical settings. Numeric covariates can be handled by the identity transformation $g_{ji}(x) = x$ with $p_j = 1$. Nominal covariates at levels $1, . . . , K$ are represented by $g_{ji}(k) = e_K(k)$, the unit vector of length $K$ with $k$th element being equal to one, and then $p_j = K$. For censored regression, the influence function $h$ may be chosen as logrank scores taking censoring into account, in which case $q=1$. In their extension of ctree to IC ctree, \citet{ICtree} specified the influence function $h$ to be the log-rank score for interval-censored data proposed by \citet{rankscore}. This score assigns a univariate scalar value $U_i$ to the bivariate response $\mathbf{Y}_i=(L_i,R_i]$, where $L_i$ and $R_i$ are the left and right endpoints of the censoring interval for the $i$-th observation. It is defined as
\begin{equation*}
U_i = \dfrac{\widehat{S}(L_i)\log\widehat{S}(L_i)-\widehat{S}(R_i)\log\widehat{S}(R_i)}{\widehat{S}(L_i)-\widehat{S}(R_i)}, \;\;\;\text{when }L_i<R_i
\end{equation*}
and 
\begin{equation*}
U_i = 1+ \log\widehat{S}(L_i) , \;\;\;\text{when }L_i=R_i,
\end{equation*}
where $\widehat{S}(\cdot)$ is the nonparametric maximum likelihood estimator (NPMLE) of the survival function. We similarly use the log-rank score $U_i$ in our proposed extension of cforest to IC cforest.

The aggregation scheme of the cforest is different from that of the random survival forest. Instead of averaging predictions directly as in the random survival forest, it works by averaging observation weights extracted from each of the individual trees and estimates the conditional survival probability function by computing one single Kaplan-Meier curve based on weighted observations identified by the leaves of bootstrap survival trees. The idea of averaging weights instead of predictions is advocated in \citet{quantile} for quantile regression. \citet{GRF} also adopt the same scheme for more general settings and propose the generalized random forest. These weights can be viewed as \q{adaptive nearest neighbor weights,} a term borrowed from \citet{adaptive}, where these weights were theoretically studied for the estimation of conditional means for regression forests. The core idea is to obtain a \q{distance} or a \q{similarity} measure based on the number of times a pair of observations is assigned to the same terminal node in the different trees of the forest. For conditional mean estimation, the averaging and weighting views of forests are equivalent; however, if we move to more general settings like constructing a nonparametric method for complex data situations, the weighting scheme has been proved to be more efficient \citep{GRF}.

Consider cforest where a set of $B$ trees is grown, indexed by $b=1,2,...,B$. Each leaf of a tree corresponds to a rectangular subspace of $\mathcal{X}$. For any new observation $\mathbf{x}\in\mathcal{X}$, for each tree there is one and only one leaf such that $\mathbf{x}$ falls into it. Denote the corresponding rectangular subspace of this leaf in the $b$-th tree as $R_b(\mathbf{x})\subseteq\mathcal{X}$. The weight of each observation $\mathbf{X}_i=(X_{1i},...,X_{mi})^T$ in the original sample, $v_{i,b}(\mathbf{x})$, measures the \q{similarity} of the $i$-th observation $\mathbf{X}_i$ to the new observed value $\mathbf{x}$ by counting how many times the value of $\mathbf{X}_i$ in the original sample falls into the same leaf as $\mathbf{x}$ in the $b$-th tree
\begin{align*}
v_{i,b}(\mathbf{x})=\dfrac{\mathbf{1}_{\{\mathbf{X}_i\in R_b(\mathbf{x})\}}}{\#\{j:\mathbf{X}_j\in R_b(\mathbf{x})\}}.
\end{align*}
Averaging over $B$ trees, the weights are
\begin{align*}
v_i(\mathbf{x})=\dfrac{1}{B}\sum_{b=1}^Bv_{i,b}(\mathbf{x}),
\end{align*}
which sum to one. The survival function can then be constructed by using a weighted version of the non-parametric maximum likelihood estimator (NPMLE). Since the weights can be viewed as replications of the corresponding observations, the corresponding log likelihood function to be maximized can be written as
\begin{align*}
\log L(S(\cdot)|\mathcal{Y},\mathbf{x})=\log\left[\prod_{i=1}^n\mathbb{P}(L_i < T_i \le R_i)^{v_i(\mathbf{x})}\right].
\end{align*}
In practice, such an estimator can be constructed using the algorithm proposed by \citet{Turnbull}. Denote the Turnbull intervals as $\mathcal{I}=\{(\tau_{11},\tau_{12}],(\tau_{21},\tau_{22}],...,(\tau_{l1},\tau_{l2}]\}$ and the mass that is assigned to $(\tau_{j1},\tau_{j2}]$ as $u_j=\mathbb{P}(\tau_{j1}<T\le\tau_{j2})=S(\tau_{j1})-S(\tau_{j2})$, for $j=1,2,...,l$. Maximization of $\log L(S(\cdot)|\mathcal{Y},\mathbf{x})$ reduces to maximization of the following log likelihood function:
\begin{align}\label{eq:llk}
\log L_T(u_1,...,u_l|\mathbf{x})=\log\left[\prod_{i=1}^n\left(\sum_{j=1}^l \alpha_j^iu_j\right)^{v_i(\mathbf{x})}\right]=\sum_{i=1}^nv_i(\mathbf{x})\left(\log\sum_{j=1}^l\alpha_j^iu_j\right),
\end{align}
where $\alpha_j^i=\mathbb{I}\{(\tau_{j-1},\tau_j]\subseteq(L_i,R_i]\}$ and the parameters are subject to the constraints $u_j\ge 0$ and $\sum_{j=1}^lu_j=1$. 
Since the weights $v_1(\mathbf{x}),...,v_n(\mathbf{x})$ define the forest-based adaptive neighborhood of $\mathbf{x}$, the resulting estimator from the weighting scheme can be viewed as a locally adaptive maximum likelihood estimator. 

The weighted version of Turnbull's self-consistent estimator of $(u_1,u_2,...,u_l)$ can be obtained as the solution of the simultaneous equation
\begin{align*}
\widehat{u}_j(\mathbf{x})=\dfrac{1}{n}\sum_{i=1}^nv_i(\mathbf{x})\dfrac{\alpha_j^i}{\sum_{k=1}^l\alpha_k^i\widehat{u}_k}\widehat{u}_j(\mathbf{x}),\;\;\;1\le j\le l.
\end{align*}
Turnbull's estimator uses a self-consistency argument to motivate an iterative algorithm for the NPMLE, which turns out to be a special case of the EM-algorithm. \citet{icnp} recently proposed an efficient implementation of the EMICM algorithm to fit the NPMLE, which greatly improves the computation power and therefore enables efficient prediction from the forest for interval-censored data. In the case of weighted observations, the EM step uses the same log likelihood function as in (\ref{eq:llk}), and the ICM step, which reparameterizes the problem in terms of the vector $\Lambda_{jk}=\log(-\log S(\tau_{jk}))$ for $k=1,2$, $j=1,2,...,l$, is to update the likelihood function as
\begin{align*}
\sum_{i=1}^nv_i(\mathbf{x})\left(\log\sum_{j=1}^l\alpha_j^i[\exp(-\exp(\Lambda_{j1}))-\exp(-\exp(\Lambda_{j2}))]\right).
\end{align*}
This is then approximated with a second-order Taylor expansion for maximization \citep{icnp}.

\subsection{Regulating the construction of the IC ctrees in the IC cforest}\label{sec:OOBdes}
As discussed in Section \ref{sec:excf} only a random subset of covariates is considered for splitting at each node. The size of this random set is denoted by \textit{mtry}. It will be shown later that \textit{mtry} is a very important tuning parameter. Other parameters such as \textit{minsplit} (the minimum sum of weights in a node in order to be considered for splitting), \textit{minprob} (the minimum proportion of observations needed to establish a terminal node) and \textit{minbucket} (the minimum sum of weights in a terminal node), which control whether or not to implement a split (and thereby regulate the size of the individual trees), can potentially be essential in avoiding overfitting, and therefore may improve the overall performance. 

The recommended values for these parameters are usually given as defaults to the algorithm. For example, \textit{mtry} is usually set to be $\sqrt{m}$, where $m$ is the number of covariates \citep{SE,RSF}. However, in practice, we find that the choice of these parameters has a non-negligible effect on the overall performance of the proposed ensemble method. \citet{book} suggests that the best values for these parameters depend on the problem and they should be treated as tuning parameters. How these parameters affect the performance of proposed IC cforest and further guidelines on how to set these values are discussed in Section \ref{sec:tuning}.

\subsection{Other ensemble resampling methods}
Recently, two papers introduced novel approaches to constructing ensemble methods for survival data. \citet{CUTs} proposed censoring unbiased regression survival trees and ensembles by extending the theory of censoring unbiased transformations applicable to loss functions for right-censored survival data. This new class of ensemble algorithms extends the random survival forest algorithm for use with an arbitrary loss function and allows the use of more general bootstrap procedures, such as the exchangeably weighted bootstrap \citep{bayesbootstrap}. The extension of the theory of censoring unbiased transformation is not applicable in our context since the conditional inference framework uses multiple testing procedures that measure the association between responses and covariates for variable selection and splitting procedure, rather than loss minimization. The exchangeably weighted bootstrap procedures, including Bayesian bootstrap and the iid weighted bootstrap with weights simulated from a Gamma distribution, assign strictly positive real-valued weights to each observation in every bootstrap sample. This is in contrast to the nonparametric bootstrap approach that is used in the conditional inference forest framework, which places positive integer weights that sum to the sample size on approximately 63\% of the observations in any given bootstrap sample. With these weights the exchangeably weighted bootstrap can avoid generating additional ties in the response variable when it is applied to censored survival data. Unfortunately, it is computationally infeasible in the conditional inference forest framework because resampling using the real-valued weights would require algorithm weights that effectively make the sample size orders of magnitude larger.

\citet{RSFSE} developed a random survival forest with space extension algorithm by combining random subspace, bagging, and extended space techniques. The extended covariate space used for model building contains all of the original covariates plus new covariates formed by differencing two randomly selected original ones. It can be applied to aggregation schemes that average predictions, as is done in the random survival forest, but is inapplicable to aggregation schemes that average observation weights, as is done in the conditional inference forest. This is because when using extended space techniques the covariates of each observation change for each bootstrap base learner replication. For these reasons we will discuss only the standard conditional inference forest construction in this paper.

\section{Properties of the conditional inference forest method}
In this section, we use computer simulations to investigate the properties of the proposed IC cforest estimation method. The event time $T$ is generated from distribution $F(t)$ and the gap $\delta_t$ between any two consecutive examination times from a distribution $G(t)$. The $j$-th of in total $k+1$ examination times therefore is $t_j = \sum_{i=1}^j \delta_i$ and the intervals will be $(0,t_1],(t_1,t_2],...,(t_k,\infty)$, each with width $\delta_i, i = 1,...,k+1$. The censoring interval of $T$ is the one that contains $T$. Here $F(t)$ and $G(t)$ are independent, and therefore the survival times $T$ and the censoring mechanism are independent. This mechanism ensures the possibility that some observations can potentially be right-censored, i.e. $T$ lies in $(t_k, \infty)$. 

We will study the properties of the proposed cforest method in terms of its estimation performance. The simulation setups are similar to those in \citet{ICtree}. 
\subsection{Model setup}
We use three simulation setups, each with five distributions ($F(t)$) of survival (event) time $T$ to test the prediction performance of the proposed IC cforest.

\begin{figure}[t!]
\begin{center}
\includegraphics[scale=0.5]{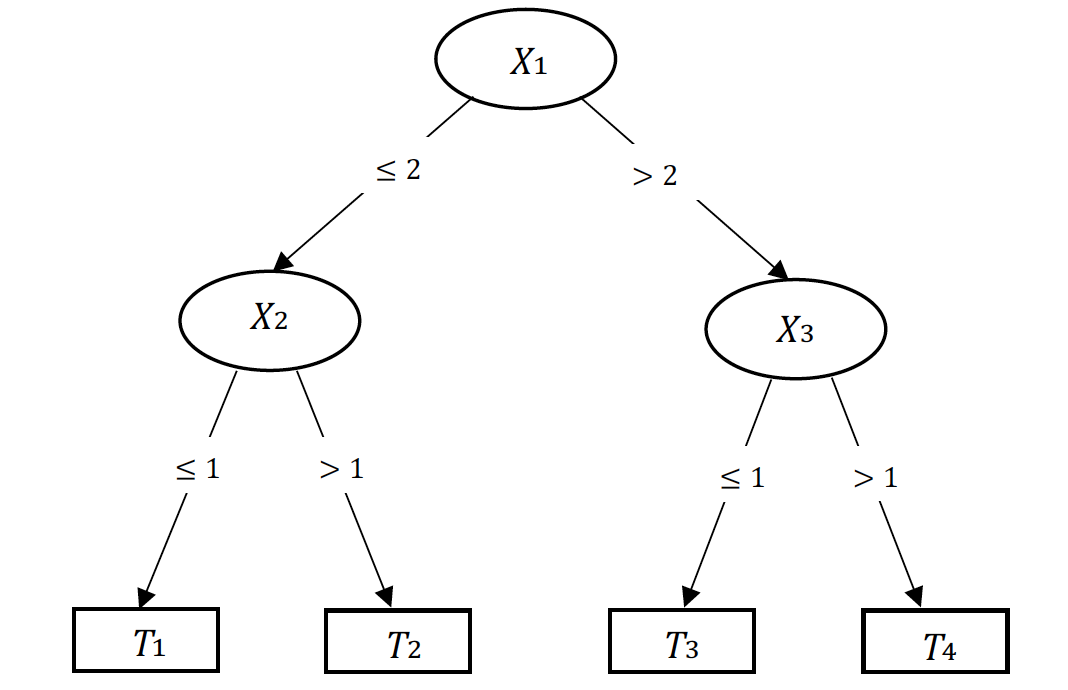}
\caption{Tree structure used in simulations.}\label{fig:tree}
\end{center}
\end{figure}

In the first setup, the underlying true model has a tree structure. There are ten covariates $X_1, ..., X_{10}$, where $X_1$, $X_4$, and $X_7$ randomly take values from the set $\{1, 2, 3, 4, 5\}$, $X_2$, $X_5$, and $X_8$ are binary $\{1, 2\}$ and $X_3$, $X_6$, $X_9$, $X_{10}$ are $U[0, 2]$. Only the first three covariates $X_1, X_2, X_3$ determine the distribution of the survival (event) time $T$. The survival time $T$ follows distribution $\widetilde{T_1}$, $\widetilde{T_2}$, $\widetilde{T_3}$ or $\widetilde{T_4}$ according to the values of $X_1$, $X_2$, $X_3$ by a tree structure given in Figure \ref{fig:tree}.

The survival time $T$ is generated from one of five different possible distributions (with each of the four (pairs of) parameter values corresponding to $\widetilde{T_1}$, $\widetilde{T_2}$, $\widetilde{T_3}$ and $\widetilde{T_4}$):
\begin{enumerate}
	\item Exponential with four different values of $\lambda$ from \{0.1, 0.23, 0.4, 0.9\}.
	\item Weibull distribution with shape parameter $\alpha = 0.9$, which corresponds to decreasing hazard with time. The scale parameter $\beta$ takes the values \{7.0, 3.0, 2.5, 1.0\}.
	\item Weibull distribution with shape parameter $\alpha = 3$, which corresponds to increasing hazard with time. The scale parameter $\beta$ takes the values \{2.0, 4.3, 6.2, 10.0\}.
	\item Log-normal distribution with location parameter $\mu$ and scale parameter $\sigma$ with 4 different pairs $(\mu, \sigma)= \{(2.0, 0.3),(1.7, 0.2),(1.3, 0.3),(0.5, 0.5)\}$.
	\item Bathtub-shaped hazard model (Hjorth, 1980). The survival function is given by
	\begin{equation*}
	S(t;a,b,c) = \dfrac{\exp\left(-\dfrac{1}{2}at^2\right)}{(1 + ct)^{b/c}},
	\end{equation*}
	with $b = 1$, $c = 5$ and $a$ set to take values \{0.01, 0.15, 0.20, 0.90\}.\\
\end{enumerate}

The second and third setups are similar to those in \citet{Bagging},
\begin{itemize}
	\item Second: Linear survival relationship with $\vartheta = -X_1-X_2$.
	\item Third: Nonlinear survival relationship with $\vartheta = -\left[-\cos((X_1 +X_2)\cdot\pi)+\sqrt{X_1 +X_2}\;\right]$.
\end{itemize}
Here $\vartheta$ is a location parameter whose value is determined by covariates $X_1$ and $X_2$.  In these settings six independent covariates $X_1, ..., X_{10}$ serve as predictor variables, with $X_2, X_3, X_6, X_8, X_9$ binary \{0, 1\} and $X_1, X_4, X_5, X_7, X_{10}$ uniform $[0, 1]$. The survival time $T_i$ again depends on $\vartheta$ with five different possible distributions:
\begin{enumerate}
	\item Exponential with parameter $\lambda = e^\vartheta$;
	\item Weibull with increasing hazard, scale parameter $\lambda = 10e^\vartheta$ and shape parameter $k=2$;
	\item Weibull with decreasing hazard, scale parameter $\lambda = 5e^\vartheta$ and shape parameter $k=0.5$;
	\item Log-normal distribution with location parameter $\mu = 1.5$ and scale parameter $\sigma=e^\vartheta$;
	\item Bathtub-shaped hazard model (Hjorth, 1980). The survival function is given by
	\begin{equation*}
	S(t;a,b,c) = \dfrac{\exp\left(-\dfrac{1}{2}at^2\right)}{(1 + ct)^{b/c}},
	\end{equation*}
	with $b = 1$, $c = 5$ and $a = e^\vartheta$.
\end{enumerate}

To see how the IC cforest compares with a (semi-)parametric model and the corresponding tree model, we also include the Cox proportional hazards model implemented in the R package \textsf{icenReg} (\citealp{icenReg}) (we will refer to this as IC Cox) and the IC ctree model implemented in the R package \textsf{LTRCtrees} (\citealp{LTRC}) in the simulations for comparison. To see the amount of information loss due to interval-censoring, the oracle versions of all three models, Cox, ctree and cforest, which are fitted using the actual event time $T$, are also included as in \citeauthor{ctree} (2006$b$).

In the second setup where $\vartheta = -X_1 -X_2$ , the linear proportional hazards assumption is satisfied, so the Cox PH model should perform best. The third setup is similar to the second except that $\vartheta$ in this setup has a more complex nonlinear structure in terms of covariates, which is potentially more like a real world application. This complex structure can make the distributions of $T_i$ satisfy neither the Cox PH model nor the tree structure. 

In all three simulation setups with five distributions $F(t)$, we consider three different distributions $G(t)$ of censoring interval width $\delta_t= t_{j+1} - t_j$, 
\begin{enumerate}
	\item $G_1(t)$, Uniform distribution $U(0.15,0.35)$;
	\item $G_2(t)$, Uniform distribution $U(0.75,0.95)$;
	\item $G_3(t)$, Uniform distribution $U(1.65,1.85)$.
\end{enumerate}
Notice that censoring interval widths generated by $G_2(t)$ should be around three times wider than those generated by $G_1(t)$, and censoring interval widths generated by $G_3(t)$ should be around seven times wider than those generated by $G_1(t)$. Intuitively, as the width of the censoring interval gets wider, less information about the actual survival time is available. 

We also consider three possible right-censoring rates, $0\%$ right-censoring, light censoring with about $20\%$ observations being right-censored, and heavy
censoring with about $40\%$ observations being right-censored. 

The simulation setup is designed to investigate the extent to which estimation performance of the proposed IC cforest deteriorates with the loss of information due to widening of censoring intervals, and also due to the increasing rate of right censoring.

\subsection{Evaluation methods}
To evaluate estimation performance, the average integrated $L_2$ distance between the true and the estimated survival curves
\begin{equation}\label{eq:L2}
\dfrac{1}{n}\sum_{i = 1}^n\dfrac{1}{\max_j(T_j)}\int_{0}^{\max_j(T_j)}[\hat{S}_i(t)-S_i(t)]^2dt
\end{equation}
is used, where $T_j$ is the (actual) event time of the $j$-th observation and $\widehat{S}_i(\cdot)$ ($S_i(\cdot)$) is the estimated (true) survival function for the $i$-th observation from a particular estimator. 

\subsection{Evaluation of tuning parameters}\label{sec:tuning}
\subsubsection{\textit{mtry} as a tuning parameter}
In the cforest algorithm, a random selection of \textit{mtry} input variables is used in each node for each tree. A split is established when all of the following criteria are met: 1) the sum of the weights in the current node is larger than \textit{minsplit}, 2) a fraction of the sum of weights of more than \textit{minprob} will be contained in all daughter nodes, 3) the sum of the weights in all daughter nodes exceeds \textit{minbucket}, and 4) the depth of the tree is smaller than \textit{maxdepth}. Default values of \textit{mtry}, \textit{minsplit},\textit{minprob}, \textit{minbucket} and \textit{maxdepth} have been given in $\mathit{ctree\_control}$ of the R package \textsf{partykit} \citep{partykit}, where \textit{mtry} is set to be $\sqrt{m}$ (where $m$ is the number of covariates), and the other four parameters are set to be $\{20,14,7,\text{Inf}\}$. Since typically unstopped and unpruned trees are used in random forests, we do not see \textit{maxdepth} as a tuning parameter in the proposed IC cforest method.

The value of \textit{mtry} can be fined-tuned on the \q{out-of-bag observations.} The \q{out-of-bag observations} for the $b$-th tree are those observations that are left out of the $b$-th bootstrap sample and not used in the construction of the $b$-th tree (in fact, about one-third of the observations in the original sample are \q{out-of-bag observations} for each bootstrap sample). The response for the $i$-th observation can then be predicted by using each of the $B$ trees in which that observation was \q{out-of-bag} (this will yield around $B/3$ predictions for the $i$-th observation). The resulting prediction error is a valid estimate of the test error for the ensemble method. The idea of tuning \textit{mtry} on the out-of-bag observations is borrowed from the function \texttt{tuneRF()} in the R package \textsf{randomForest} \citep{randomForest}. A version of \texttt{tuneRF()} for interval-censored data starts with the default values of \textit{mtry}, then searches for the optimal values with a prespecified step factor with respect to out-of-bag error estimate \textit{mtry} for IC cforest. The integrated Brier score \citep{brier}, which is the most popular measure of prediction error in survival analysis, is used in the function \texttt{tuneRF()} for right-censored time data. \citet{icbs} adapted the integrated Brier score to interval-censored time data,
\begin{equation}\label{eq:ibs}
\dfrac{1}{n}\sum_{i = 1}^n\dfrac{1}{T_{\mathrm{max}}}\int_{0}^{T_{\mathrm{max}}}[\mathbb{I}(T_i>t)-\widehat{S}_i(t)]^2dt
\end{equation}
with $T_{\mathrm{max}}=\max_{i=1,...,n}\{L_i,R_i\}$ and $\mathbb{I}(T_i>t)$ estimated by 
\begin{equation*}
\widehat{\mathbb{I}}(T_i>t) = \dfrac{\widehat{S}_i(t)-\widehat{S}_i(R_i)}{\widehat{S}_i(L_i)-\widehat{S}_i(R_i)},
\end{equation*}
where $\widehat{S}_i(\cdot)$ is the estimated survival function for the $i$-th observation. Using this evaluation measure we can tune the \textit{mtry} by the \q{out-of-bag} tuning procedure given in Appendix \ref{sec:oobmtry}.
\begin{figure}[t!]
\includegraphics[scale=0.67]{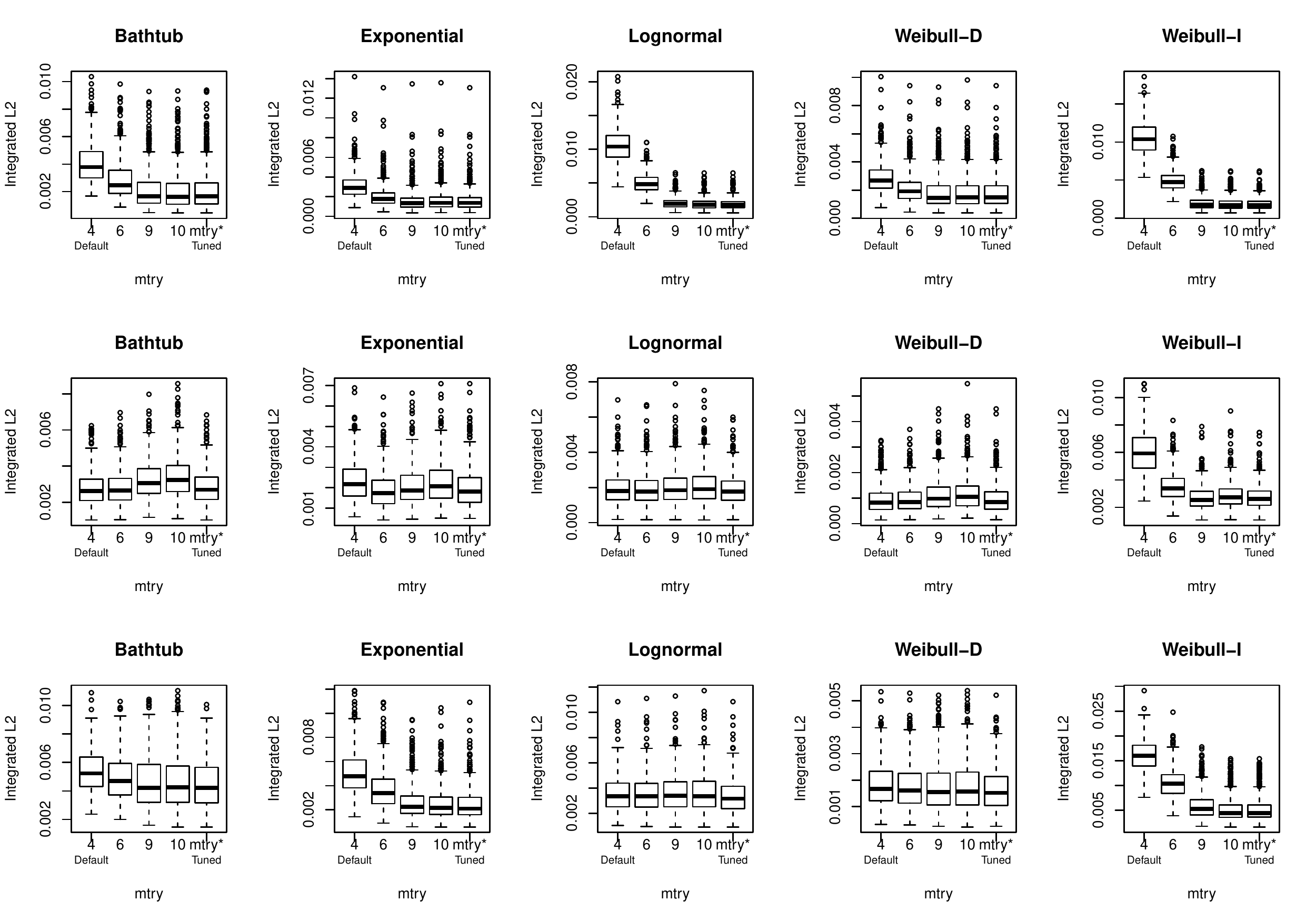}
\caption{\small{Integrated $L_2$ difference of IC cforest with different \textit{mtry} values, with $n = 200$, no right censoring and the interval censoring width generated by $G_1(t)$. The default value in \texttt{cforest} function is $\sqrt{m} \approx 4$. The value of \textit{mtry} tuned by the \q{out-of-bag} tuning procedure is given in the last column in each boxplot. Top row gives results for the first setup (tree structure), middle row gives results for the second setup (linear model), and bottom row gives results for the third setup (nonlinear model).}}\label{fig:mtry}
\end{figure}

Figure \ref{fig:mtry} gives an example of how IC cforest performs with different values of \textit{mtry}. The \textit{mtry} values are chosen using \textit{stepFactor} $s=1.5$ in the algorithm given in Appendix \ref{sec:oobmtry}.  In this example, the default value of \textit{mtry} in the \texttt{cforest} function is not always optimal and sometimes the performance can be significantly improved by setting a larger value (values smaller than the default value never had better performance, so they are not given). In fact, different distributions with different underlying models favor different values of \textit{mtry}. The \q{out-of-bag} tuning procedure provides a relatively reliable choice of \textit{mtry} that gives relatively good performance overall.

The size $n = 200$ with no right censoring and the censoring interval width generated by $G_1(t)$ is used in the simulations presented in Figure \ref{fig:mtry}; results with $n = 500$ and $n = 1000$ were similar and are given in Appendix \ref{sec:etN51} and Appendix \ref{sec:etN11}. 
\subsubsection{\textit{minsplit}, \textit{minprob} and \textit{minbucket} as tuning parameters}
The optimal values that determine the split vary from case to case. As a fixed number, the default values may not affect the splitting at all when the sample size is large, while having a noticeable effect in smaller data sets. This inconsistency can potentially result in good performance in some data sets and poor performance in others. Here we wish to determine a rule that can automatically adjust those values to the size of the data set, whose performance is relatively stable and better than that of the default values. 

The values of \textit{minsplit}, \textit{minprob} and \textit{minbucket} determine whether a split in a node will be implemented. We design our experiments to explore the individual effect of each parameter. Based on the results, we propose the \q{15\%-Default-6\% Rule,} which is to set \textit{minsplit} to be 15\% of the sample size $n$, \textit{minprob} to be the default value, and \textit{minbucket} to be 6\% of the sample size $n$.

\begin{figure}[t!]
\includegraphics[scale=0.67]{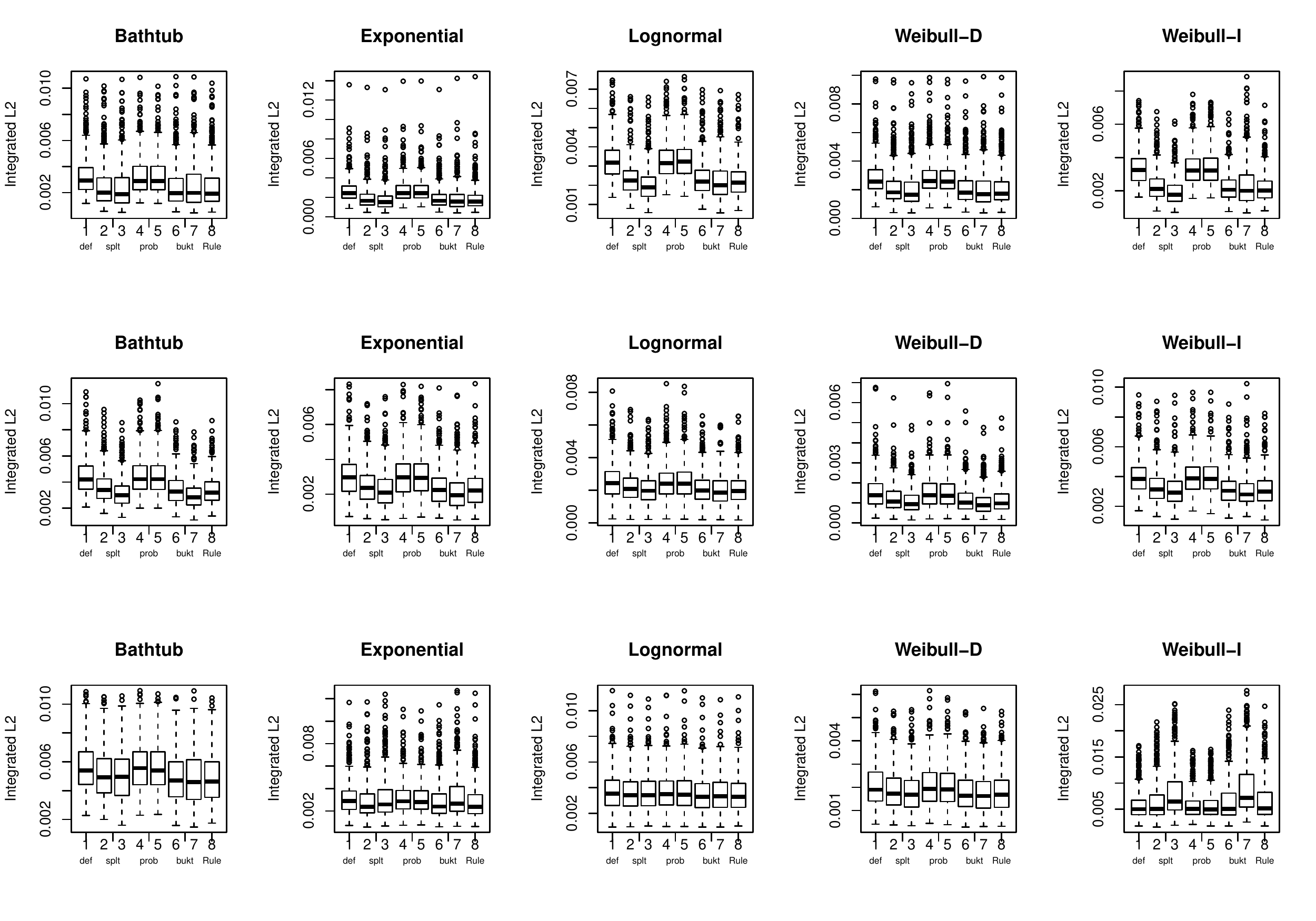}
\caption{\small{Example: integrated $L_2$ difference of IC cforest with different \textit{minsplit}, \textit{minprob} and \textit{minbucket} values, with $n = 200$, no right-censoring and the interval-censoring width generated by $G_1(t)$. 1-(\textit{minsplit}, \textit{minprob}, \textit{minbucket}) $=(20,0.01,7)$. 2-(\textit{minsplit}, \textit{minprob}, \textit{minbucket}) $=(30,0.01,7)$, 3-(\textit{minsplit}, \textit{minprob}, \textit{minbucket}) $=(40,0.01,7)$, 4-(\textit{minsplit}, \textit{minprob}, \textit{minbucket}) $=(20,0.05,7)$, 5-(\textit{minsplit}, \textit{minprob}, \textit{minbucket}) $=(20,0.10,7)$, 6-(\textit{minsplit}, \textit{minprob}, \textit{minbucket}) $=(20,0.01,12)$, 7-(\textit{minsplit}, \textit{minprob}, \textit{minbucket}) $=(20,0.01,16)$. 8-The \q{15\%-Default-6\% Rule}: (\textit{minsplit}, \textit{minprob}, \textit{minbucket}) $=(30,0.01,12)$. Top row gives results for the first setup (tree structure), middle row gives results for the second setup (linear model), and bottom row gives results for the third setup (nonlinear model).}} \label{fig:3param}
\end{figure}

Figure \ref{fig:3param} gives an example of the sensitivity of IC cforest to the different values of \textit{minsplit}, \textit{minprob}, and \textit{minbucket}. The choices of \textit{minsplit} are 20 (default value), 30 (15\% of the sample size $n$), and 40 (20\% of the sample size $n$). The choices of \textit{minprob} are 0.01 (default value), 0.05, and 0.10. The choices of \textit{minbucket} are 7 (default value), 12 (6\% of the sample size $n$), and 16 (8\% of the sample size $n$). In each plot of Figure \ref{fig:3param}, column 1 shows the integrated $L_2$ under the default setting, columns 2-7 show the the integrated $L_2$ differences when changing the value of one parameter at a time while holding the others the same, and column 8 shows the results of the proposed \q{15\%-Default-6\% Rule.} Here the performance of IC cforest is shown with a limited number of values and these values are selected to give as much understanding of the performance change due to the tuning parameters as possible. We can see that overall the value of \textit{minprob} does not change the performance much (as expected, since we set the equivalent parameter, \textit{minbucket}, to be a much larger proportion of the size of the data set), while changing \textit{minsplit} and \textit{minbucket} can possibly improve the performance of the overall performance. Empirically, the \q{15\%-Default-6\% Rule} has shown to improve the overall performance over the default setting under different models with different distributions. The simulation results show that a slightly larger size of leaf is favored, since the smaller default size makes the forest more prone to capturing noise and overfitting, and therefore exhibits worse performance.

The size $n = 200$ with no right censoring and the censoring interval width generated by $G_1(t)$ is used in the simulations presented here; results with $n = 500$ and $n = 1000$ were similar and are given in Appendix \ref{sec:etN52} and Appendix \ref{sec:etN12}. 
\subsection{Estimation performance}
We run 500 simulation trials for each setting to see how well the proposed IC cforest performs compared to the IC Cox model and the corresponding IC ctree model. The parameter \textit{mtry} in IC cforest is tuned following the \q{out-of-bag} tuning procedure and the values for \textit{minsplit}, \textit{minprob} and \textit{minbucket} are chosen using the \q{15\%-Default-6\% Rule} described in Section \ref{sec:tuning}. The size $n = 200$ with censoring interval width generated by $G_1(t)$ is used in the simulations presented here; results with $n = 500$ and $n = 1000$ were similar and are given in Appendix \ref{sec:epN5} and Appendix \ref{sec:epN1}, respectively. 

\begin{figure}[t!]
\includegraphics[scale=0.67]{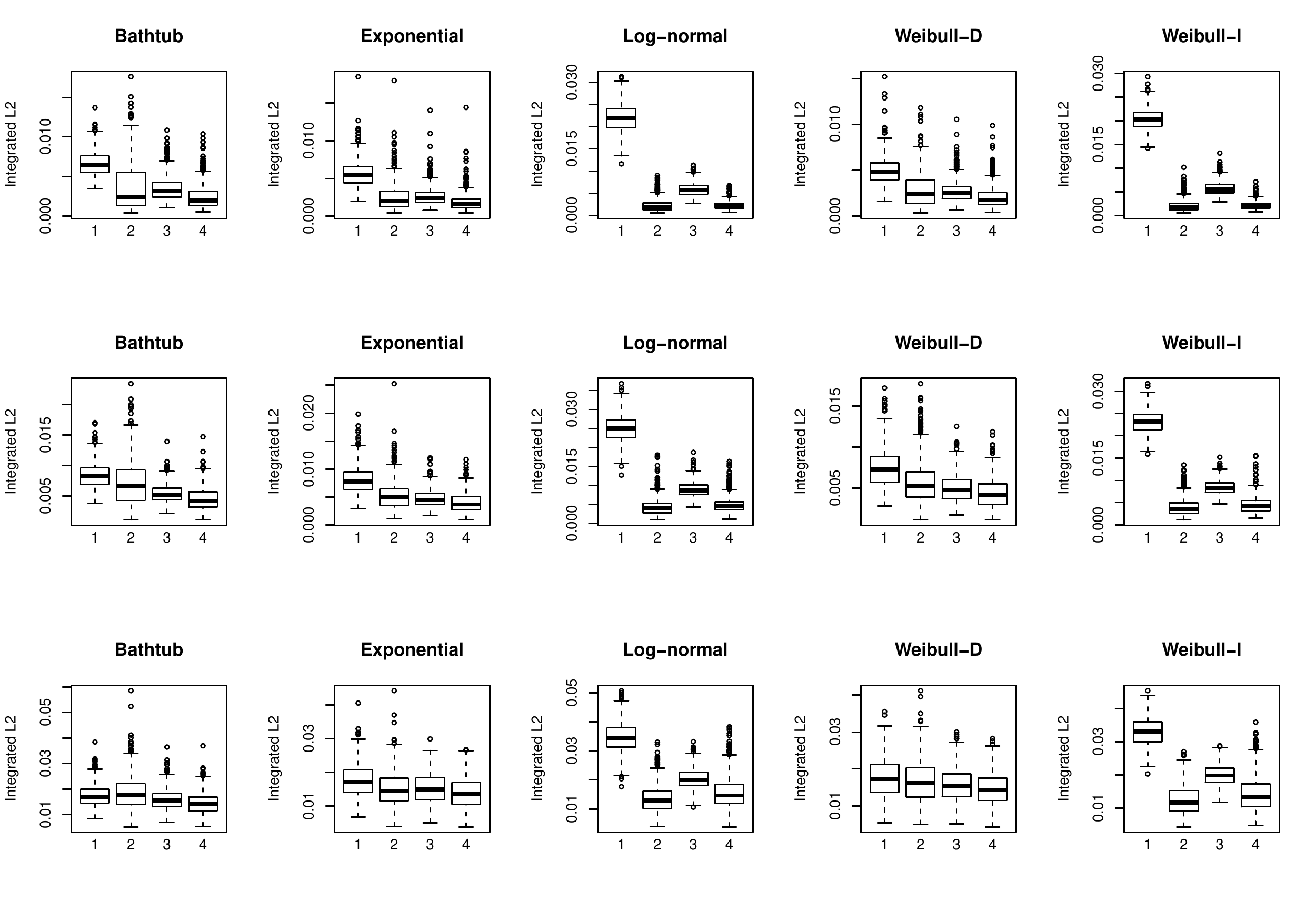}
\caption{\small{True tree model with censoring interval width generated from $G_1(t)$: integrated $L_2$ difference boxplots with $n=200$. Methods are numbered as 1-IC Cox model, 2-IC ctree, 3-IC cforest with parameters set by default, 4-IC cforest with parameters set through \q{out-of-bag} tuning procedure and the \q{15\%-Default-6\% Rule.} Top row gives results without right-censoring, middle row gives results for light (right-)censoring, and bottom row gives results for heavy (right-)censoring.}}\label{fig:w1m1}
\end{figure}

\begin{figure}[t!]
\includegraphics[scale=0.65]{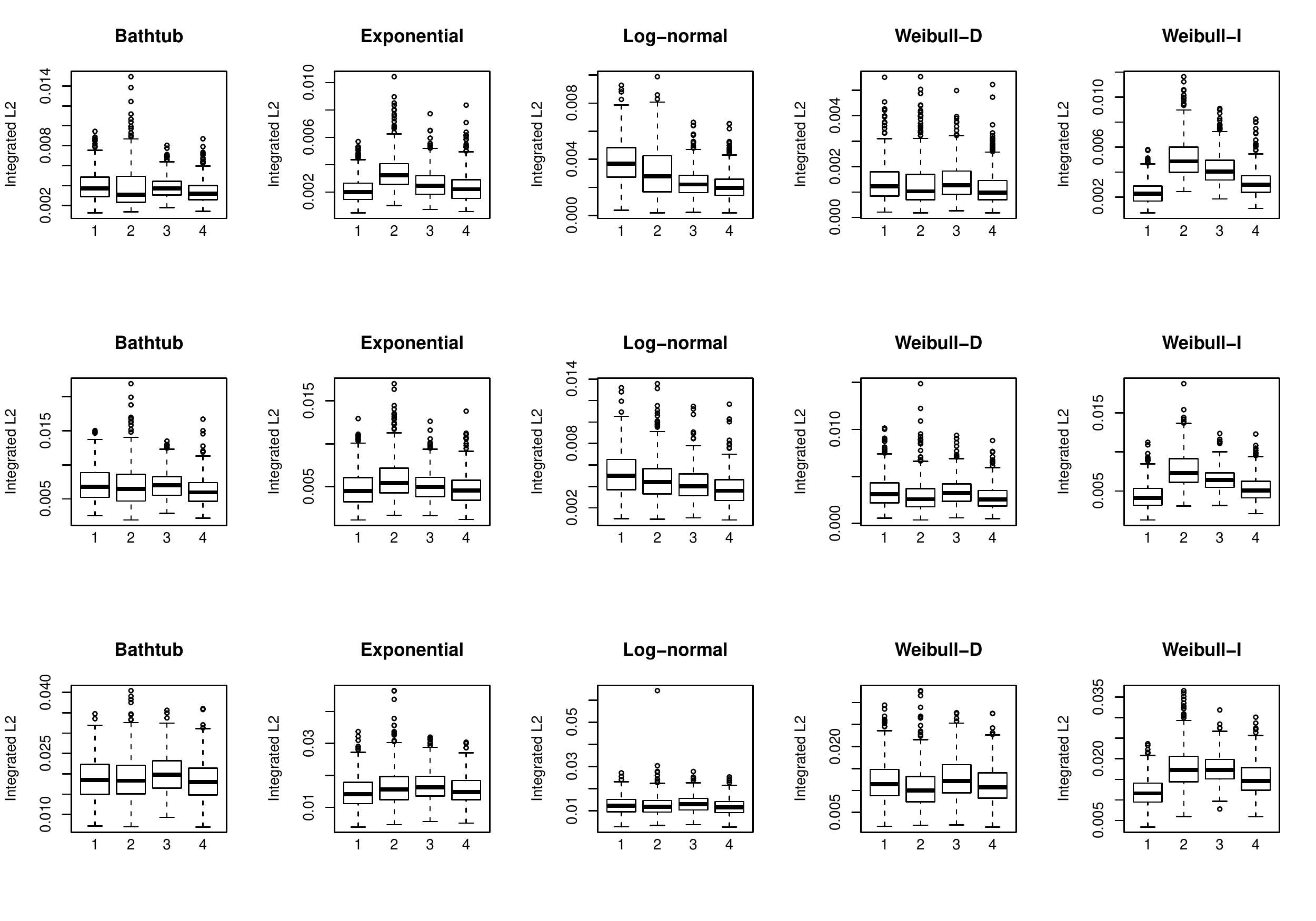}
\caption{\small{True linear model with censoring interval width generated from $G_1(t)$: integrated $L_2$ difference boxplots with $n=200$. Methods are numbered as 1-IC Cox model, 2-IC ctree, 3-IC cforest with parameters set by default, 4-IC cforest with parameters set through \q{out-of-bag} tuning procedure and the \q{15\%-Default-6\% Rule.} Top row gives results without right-censoring, middle row gives results for light (right-)censoring, and bottom row gives results for heavy (right-)censoring.}}\label{fig:w1m2}
\end{figure}

\begin{figure}[t!]
\includegraphics[scale=0.65]{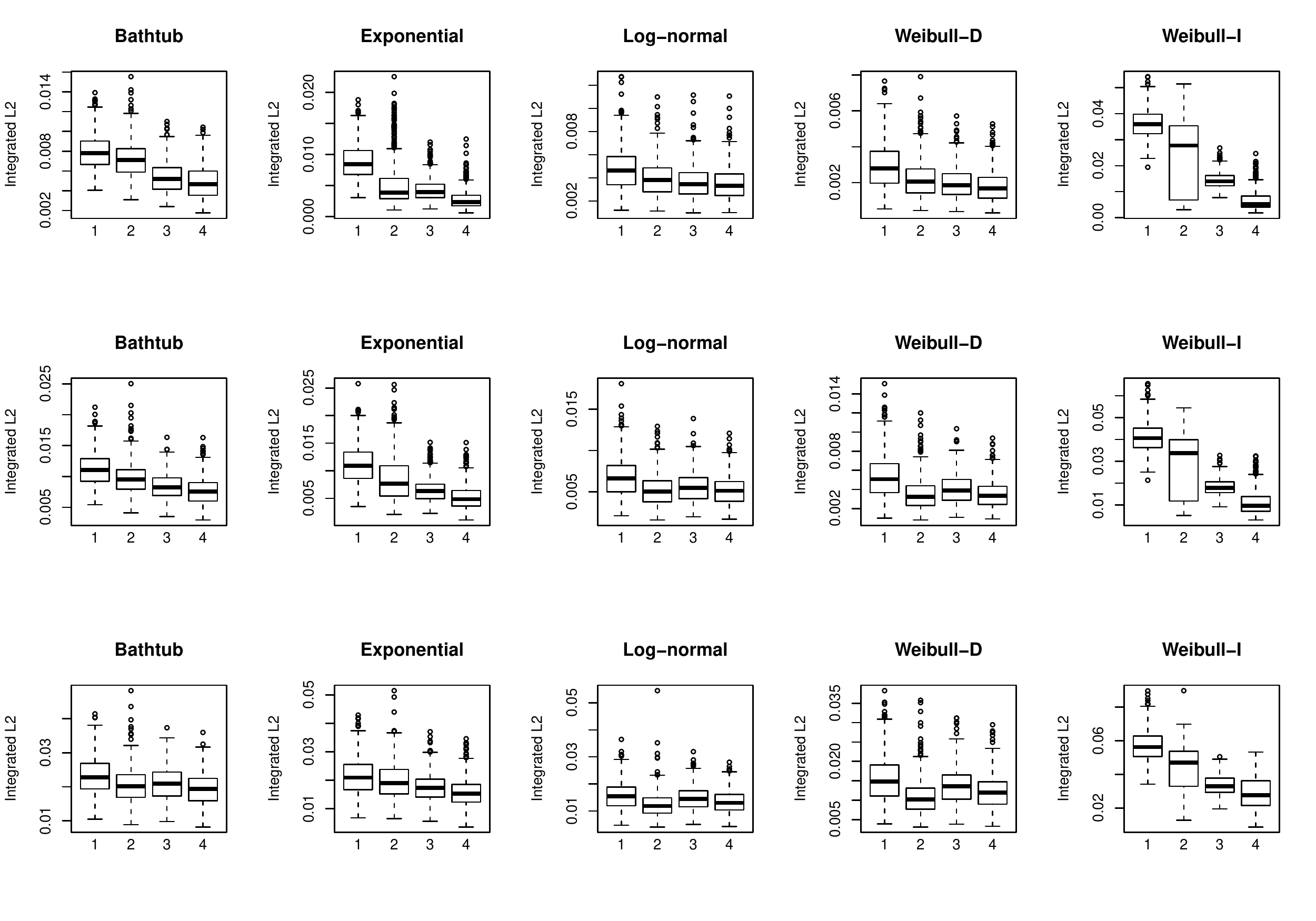}
\caption{\small{True nonlinear model with censoring interval width generated from $G_1(t)$: integrated $L_2$ difference boxplots with $n=200$. Methods are numbered as 1-IC Cox model, 2-IC ctree, 3-IC cforest with parameters set by default, 4-IC cforest with parameters set through \q{out-of-bag} tuning procedure and the \q{15\%-Default-6\% Rule.} Top row gives results without right-censoring, middle row gives results for light (right-)censoring, and bottom row gives results for heavy (right-)censoring.}}\label{fig:w1m3}
\end{figure}

Figures \ref{fig:w1m1} to \ref{fig:w1m3} give side-by-side integrated $L_2$ difference boxplots for all three setups with sample size $n=200$ with censoring width generated from $G_1(t)$. We can see that the \q{out-of-bag} tuning procedure and the \q{15\%-Default-6\% Rule} improve the IC cforest performance over the parameters set by default. Figure \ref{fig:w1m1} shows that in the presence of right-censoring, the proposed IC cforest performs as least as well as the IC ctree method in the first setup, where the true model is a tree. In addition, for all five distributions, the IC cforest outperforms the IC Cox model. 

As expected, the IC Cox model can outperform the IC cforest method in the second setup (where the true model is a linear model). This occurs when the underlying distribution is the Weibull-Increasing distribution, but for other distributions and up to right-censoring rate $40\%$, the proposed IC cforest can represent a linear model as well as the IC Cox model or even better than it. 

IC ctree outperforms IC Cox model in the third setup due to its flexible structure \citep{ICtree}, and we can see in Figure \ref{fig:w1m3}, the proposed IC cforest further improves the performance and shows its advantage in a relatively complex survival relationship. 

The censoring interval width generating distribution $G_1(t)$ is used in the simulations presented here. Intuitively, a wider censoring interval, meaning less information and more uncertainty, will result in poorer performance in the forest. 
\begin{figure}[t!]
\includegraphics[scale=0.65]{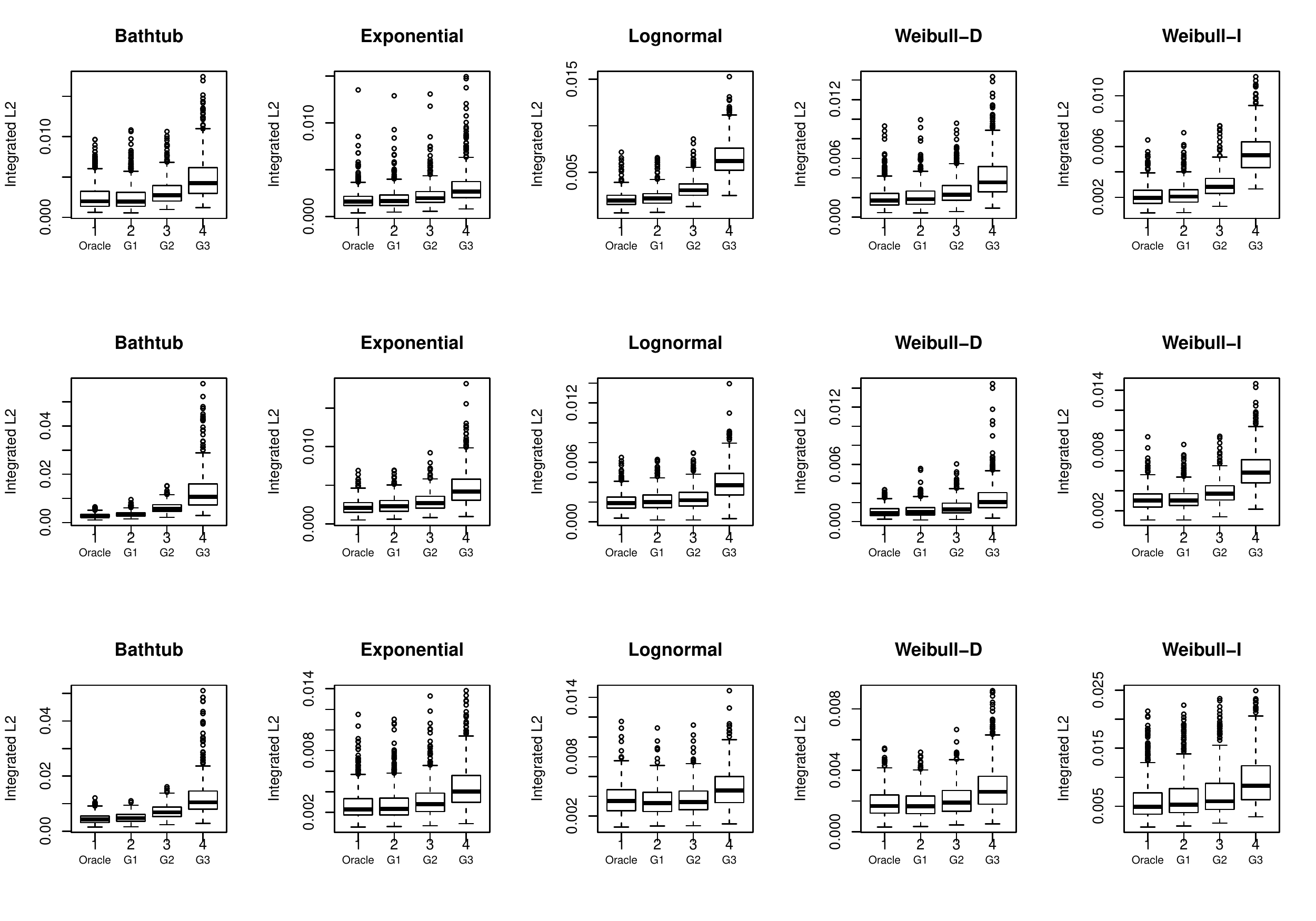}
\caption{\small{Integrated $L_2$ difference boxplots with $n=200$, no right-censoring. 1-Oracle, 2-censoring interval width generated from $G_1(t)$, 3-Censoring interval width generated from $G_2(t)$, 4-Censoring interval width generated from $G_3(t)$. Methods that give results in columns 2-4 are IC cforest with \textit{mtry} chosen through \q{out-of-bag} tuning procedure and \textit{minsplit}, \textit{minprob}, \textit{minbucket} chosen following \q{15\%-def-6\% Rule.} Top row gives results for tree model, middle row gives results for linear model, and bottom row gives results for nonlinear model.}}\label{fig:widthcf}
\end{figure}

\begin{figure}[t!]
\includegraphics[scale=0.65]{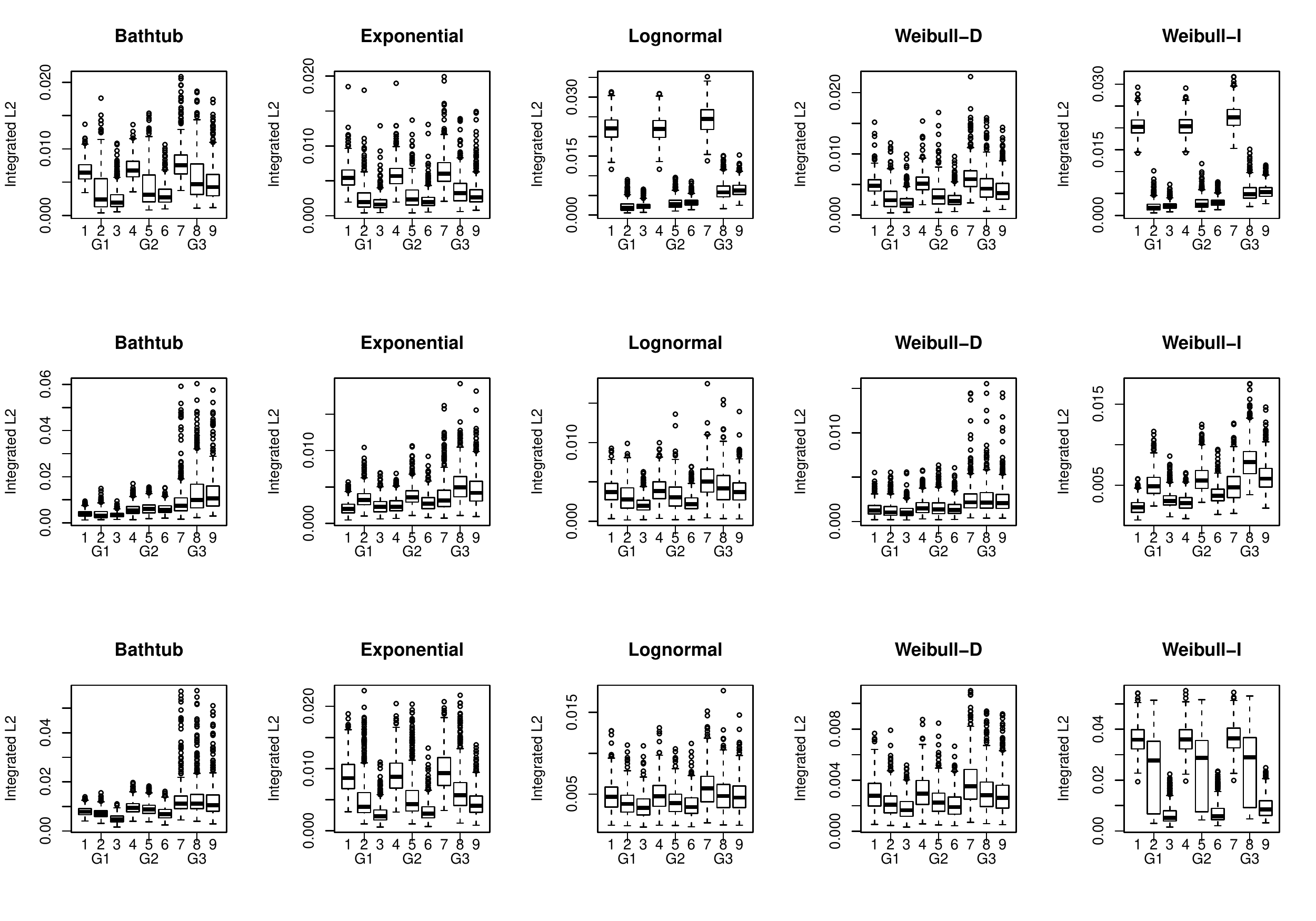}
\caption{\small{Integrated $L_2$ difference boxplots with $n=200$, no right-censoring. In each boxplot, 1-3 give results of IC Cox, IC ctree and IC cforest for censoring interval width generated from $G_1(t)$ respectively, 4-6 gives results of IC Cox, IC ctree and IC cforest for censoring interval width generated from $G_2(t)$ respectively, 7-9 give results of IC Cox, IC ctree and IC cforest for censoring interval width generated from $G_3(t)$ respectively. Top row gives results for tree model, middle row gives results for linear model, and bottom row gives results for nonlinear model.}}\label{fig:widthxtf}
\end{figure}

Figure \ref{fig:widthcf} shows how the censoring interval width affects the performance of IC cforest. When the censoring interval width is small, IC cforest can perform as well as the \q{Oracle,} where the true survival times are known, and there is no right-censoring. When the censoring interval width is roughly three times wider, the loss of information starts to affect the IC cforest performance, but not greatly. When the censoring interval width is roughly seven times wider, the IC cforest performance deteriorates considerably more. 

In fact, this loss of information due to the increased censoring interval widths affects all three different methods, and the patterns across methods we have seen in Figures \ref{fig:w1m1} to \ref{fig:w1m3} with censoring interval width generating distribution $G_1(t)$ are similar to those with $G_2(t)$ and $G_3(t)$. That is, the proposed IC cforest can still outperform the IC ctree method even under the tree model and outperform the IC Cox model under a linear model. Figure \ref{fig:widthxtf}, for example, demonstrates that the patterns across the three methods for each model preserve well under the change of censoring interval widths in the situation with no right-censoring.

\section{Real data set}
The Signal Tandmobiel\textsuperscript{\textregistered} study is a longitudinal prospective oral health study that was conducted in the Flanders region of Belgium from 1996 to 2001. In this study, 4430 first year primary school schoolchildren were randomly sampled at the beginning of the study and were dental-examined annually by trained dentists. The data consist of at most 6 dental observations for each child including time of tooth emergence, caries experience, and data on dietary and oral hygiene habits. The details of study design and research methodology can be found in \citet{datasource}. The data are provided as the tandmob2 data set in the R package \textsf{bayesSurv} \citep{bayesSurv}. The tandmob2 data set provides the time to emergence of 28 teeth in total. Each of the tooth emergence times can be taken as a response variable and we can test the prediction performance of the proposed IC cforest method, compared to the corresponding IC ctree method and IC Cox method. Potential predictors of emergence time of the child's tooth include gender, province, evidence of fluoride intake, type of educational system, starting age of brushing teeth, whether each of the twelve deciduous teeth were decayed or missing due to caries or filled, whether each of the twelve deciduous teeth were removed because of orthodontic reasons, and whether each of the twelve deciduous teeth were removed due to the orthodontic reasons or decayed on at most the last examination before the first examination when the emergence of the permanent successor was recorded. These potential predictors cover all of the variables in the data set.

To compare different methods, we conducted leave-one-out cross-validation on the entire data set, and then computed the average absolute prediction distance below $L_i$ or above $R_i$ when the predicted median emergence time falls outside of the observed interval, which measures the distance away from the interval for those observations (if a predicted emergence time falls within the observed emergence interval it is impossible to say what the prediction error is, so such observations are not considered). 
\begin{table}[t!]
\caption{Evaluation on 28 tooth data sets in Signal Tandmobiel\textsuperscript{\textregistered} Study.}\label{tl:realset}
\centering
  \begin{tabular}{ccccccc}
    \toprule
    \multirow{2}{*}{\hspace{2em}Tooth\hspace{2em}} &
      \multicolumn{2}{c}{\hspace{2.5em}IC Cox}\hspace{2.5em} &
     \multicolumn{2}{c}{\hspace{2.5em}IC ctree}\hspace{2.5em} &
     \multicolumn{2}{c}{\hspace{2.5em}IC cforest}\hspace{2.5em} \\
          \cmidrule{2-7}
      &$p_{\text{out}}(\%)$ & $\bar{d}_\text{out}$ & $p_{\text{out}}(\%)$ &$\bar{d}_\text{out}$ &$p_{\text{out}}(\%)$ &$\bar{d}_\text{out}$ \\
     \midrule
11 & 33.7 &0.3558 &33.0 &\textbf{0.3489} &32.1& 0.3732  \\
21 & 34.2&  \textbf{0.3428} & 33.2& 0.3439  &33.7& 0.3639 \\
31 & 23.6& 84.1325& 21.5&\textbf{0.3195} &20.9& 0.3312\\
41 &21.4& 71.1985& 17.4&0.6236& 18.0 & \textbf{0.6019}\\
12 &54.0 &  0.5259 &52.6& 0.5369& 54.3&  \textbf{0.5187}\\
22 &51.0& 0.5215& 50.3&  0.5232& 52.1  &\textbf{0.5026}\\
32&38.1&  0.4036& 37.4&  0.4050 &37.7 & \textbf{0.4010}\\
42&39.4& 0.4004 &38.1 & 0.4110 &39.5  &\textbf{0.3969}\\
13& 57.8&  0.6894& 57.6& \textbf{0.6236} &56.7&  0.6564\\
23& 59.1&  1.3304& 60.6&  0.5863& 60.1 & \textbf{0.5822}\\
33& 64.4& 0.6454& 71.3&  \textbf{0.6279}& 65.6& 0.6926\\
43& 63.6& 0.6386& 63.6& 0.6434& 64.6&  \textbf{0.6304}\\
14& 66.8& \textbf{0.7239} & 65.6&  0.7479 & 67.0 &  0.7311 \\
24 &67.0&  0.7082 & 68.0&\textbf{0.6934}& 66.8&  0.7176\\
 34 & 66.1 & \textbf{0.6976}& 66.4 & 0.7012& 66.3 & 0.7109\\
 44 & 65.0& 0.7108& 65.8& \textbf{0.7022} &66.6 & 0.7221\\
 15&55.6& 0.7141 &58.7&  0.6602 &56.4& \textbf{0.6382}\\
 25&55.9&  2.0519& 60.1&  0.6635 & 58.5&  \textbf{0.6629}\\
 35&52.6&  0.7245& 56.6&  0.6670 &55.9  &\textbf{0.6401}\\
45& 51.5& 0.7221& 52.4 & 0.6866& 54.7&  \textbf{0.6374}\\
16 & 25.5& \textbf{0.3138}& 22.0& 0.3765& 23.3&  0.3470\\ 
26&26.4& 0.3250& 22.8& 0.3300 & 22.8& \textbf{0.3237}\\
36 &27.5&  0.4036 & 28.0&  \textbf{0.3274}& 27.0&0.3304\\ 
46&26.6&  \textbf{0.3125}& 24.1& 0.3277&24.3& 0.3234 \\
17 & 28.8&55.2018 &28.5  &28.0678 & 28.0& \textbf{11.4780}\\
27&30.6&96.5333 & 31.3& 43.3953 & 30.9  &\textbf{30.2143}\\
37&46.3& 0.5876 &48.2&  \textbf{0.5157}& 47.2& 0.5436\\
47&43.1& 6.1757 & 46.3& \textbf{0.5615} & 43.7&0.5935\\
   \midrule[\heavyrulewidth]
   \multicolumn{7}{l}{\footnotesize$^*$ $p_{\text{out}}$-Proportion of the predicted median emergence times lying outside censoring intervals.}\\ 
    \multicolumn{7}{l}{\footnotesize$^*$ $\bar{d}_{\text{out}}$-Average absolute prediction distance below $L_i$ or above $R_i$.} \\
    \multicolumn{7}{l}{\footnotesize$^*$ The bolded value in each row indicates the smallest one among the three $\bar{d}_{\text{out}}$'s.}\\ 
  \bottomrule
  \end{tabular}
\end{table}

The IC cforest method applied with \textit{mtry} chosen through the \q{out-of-bag} tuning procedure and \textit{minplit}, \textit{minprob}, \textit{minbucket} chosen by the \q{15\%-Default-6\% Rule,} IC ctree, and the IC Cox model are applied to each of the tooth data sets. Table \ref{tl:realset} shows that the proportion of the time the predicted median emergence falls outside the observed intervals is roughly the same for the three methods, although it varies greatly from tooth to tooth. Among these 28 tooth data sets IC cforest gives the smallest average absolute prediction distance away from the observed intervals for those observations that fall outside of them for 50\% of the teeth; the IC ctree follows (32\%) and the IC Cox model trails both (18\%). Thus, the IC cforest method does a good job of predicting the actual emergence times.

\section{Conclusion}
In this paper, we have proposed a new ensemble algorithm based on the conditional inference survival forest designed to handle interval-censored data. Through the use of a simulation study, we see that the proposed IC cforest method can outperform the IC ctree and the IC Cox proportional hazards model even when the underlying true model is designed for the tree structure or the linear relationship, respectively, in terms of prediction performance, and clearly outperforms both in the nonlinear situation that neither is designed for.

The tuning parameters in the proposed IC cforest affect the overall performance of the method. In this paper, we have provided guidance on how to choose those parameters to improve on the potentially poor performance of the default settings. Further investigation of the best way to choose these parameters in a data-dependent way would be useful. It would also be interesting to extend these results to competing risks data.

An R package, \textsf{ICcforest}, that implements the IC cforest method is available at CRAN.

\section*{Acknowledgements}
Data collection of the Signal Tandmobiel$^{\textsuperscript{\textregistered}}$ data was supported by Unilever, Belgium. The Signal-Tandmobiel project comprises the following partners: Dominique Declerck (Department of Oral Health Sciences, KU Leuven), Luc Martens (Dental School, Gent Universiteit), Jackie Vanobbergen (Oral Health Promotion and Prevention, Flemish Dental Association and Dental School, Gent Universiteit), Peter Bottenberg (Dental School, Vrije Universiteit Brussel), Emmanuel Lesaffre (L-Biostat, KU Leuven), and Karel Hoppenbrouwers (Youth Health Department, KU Leuven; Flemish Association for Youth Health Care).

\

\

\newpage

\

\

\
\newpage
\bibliographystyle{abbrvnat}
\bibliography{sim}

\newpage
\begin{appendices}
\section{Algorithm of \q{out-of-bag} tuning procedure}\label{sec:oobmtry}
\begin{algorithm}[H]
	\caption{\q{Out-of-bag} tuning procedure for \textit{mtry}}
	\begin{algorithmic}[1]
		\Procedure{tuneICCF}{$\{x,L,R\}_{i=1}^n$, stepFactor}
		\State $s\gets$ stepFactor
		\State $r_1\gets$ $\min \{r\in\mathbb{N};\sqrt{m}/s^r>1\}$
		\State $r_2\gets$ $\max \{r\in\mathbb{N};\sqrt{m}s^r<m\}$
		\State $mtrypool\gets\{1,\sqrt{m}/s^{r_1},\sqrt{m}/s^{r_1-1},...,\sqrt{m}s^{r_2-1},\sqrt{m}s^{r_2},m\}$
		\For {$mtry$ in $mtrypool$}
		\State iccf.obj $\gets$ ICcforest$(\text{data} = \{x,L,R\}_{i=1}^n$, mtryTest = \textit{mtry}) 
		\State pred.oob $\gets$ predict(iccf.obj, OOB = TRUE)
		\State err.oob $\gets$ sbrier\_IC($\{x,L,R\}_{i=1}^n$, pred.oob) \Comment{calculating IBS defined in (\ref{eq:ibs})}
         \EndFor
         \State \textbf{end}
         \State $i^\ast\gets\arg\min$ err.oob
         \State $mtry^\ast\gets mtrypool[i^\ast]$
		
		\noindent	    
		\Return $mtry^\ast$.
		\EndProcedure
	\end{algorithmic}
\end{algorithm}
\

\

\newpage

\

\
\section{Evaluation of tuning parameters for $N = 500$}
\subsection{\textit{mtry} as a tuning parameter} \label{sec:etN51}
\begin{figure}[H]
\includegraphics[scale=0.67]{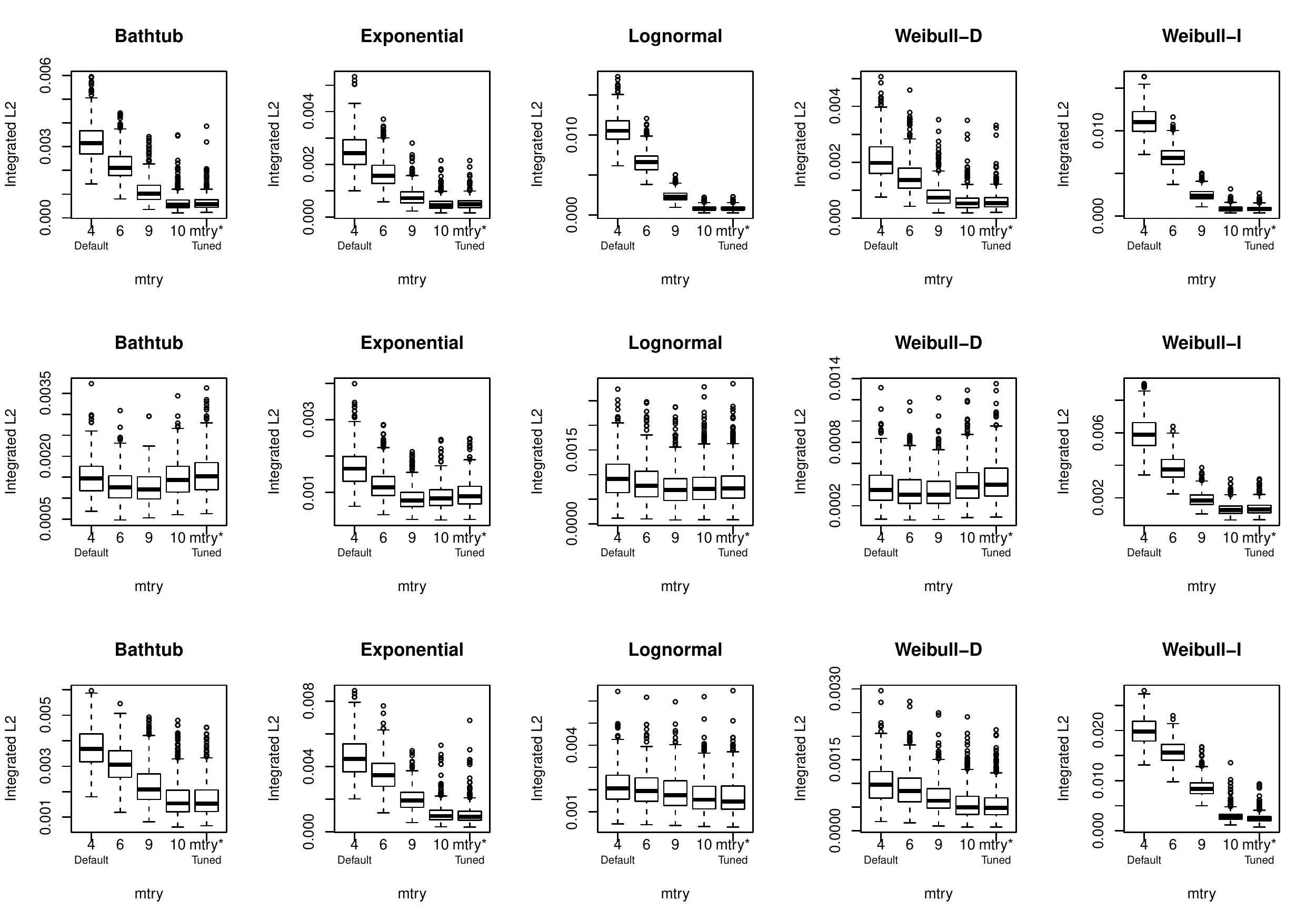}
\caption{\small{Integrated $L_2$ difference of IC cforest with different \textit{mtry} values, with $n = 500$, no right censoring and the interval censoring width generated by $G_1(t)$. The default value in \texttt{cforest} function is $\sqrt{m} \approx 4$. The value of \textit{mtry} tuned by the \q{out-of-bag} tuning procedure is given in the last column in each boxplot. Top row gives results for the first setup (tree structure), middle row gives results for the second setup (linear model), and bottom row gives results for the third setup (nonlinear model).}}
\end{figure}

\

\

\newpage

\

\

\subsection{\textit{minsplit}, \textit{minprob} and \textit{minbucket} as tuning parameters} \label{sec:etN52}
\begin{figure}[H]
\includegraphics[scale=0.67]{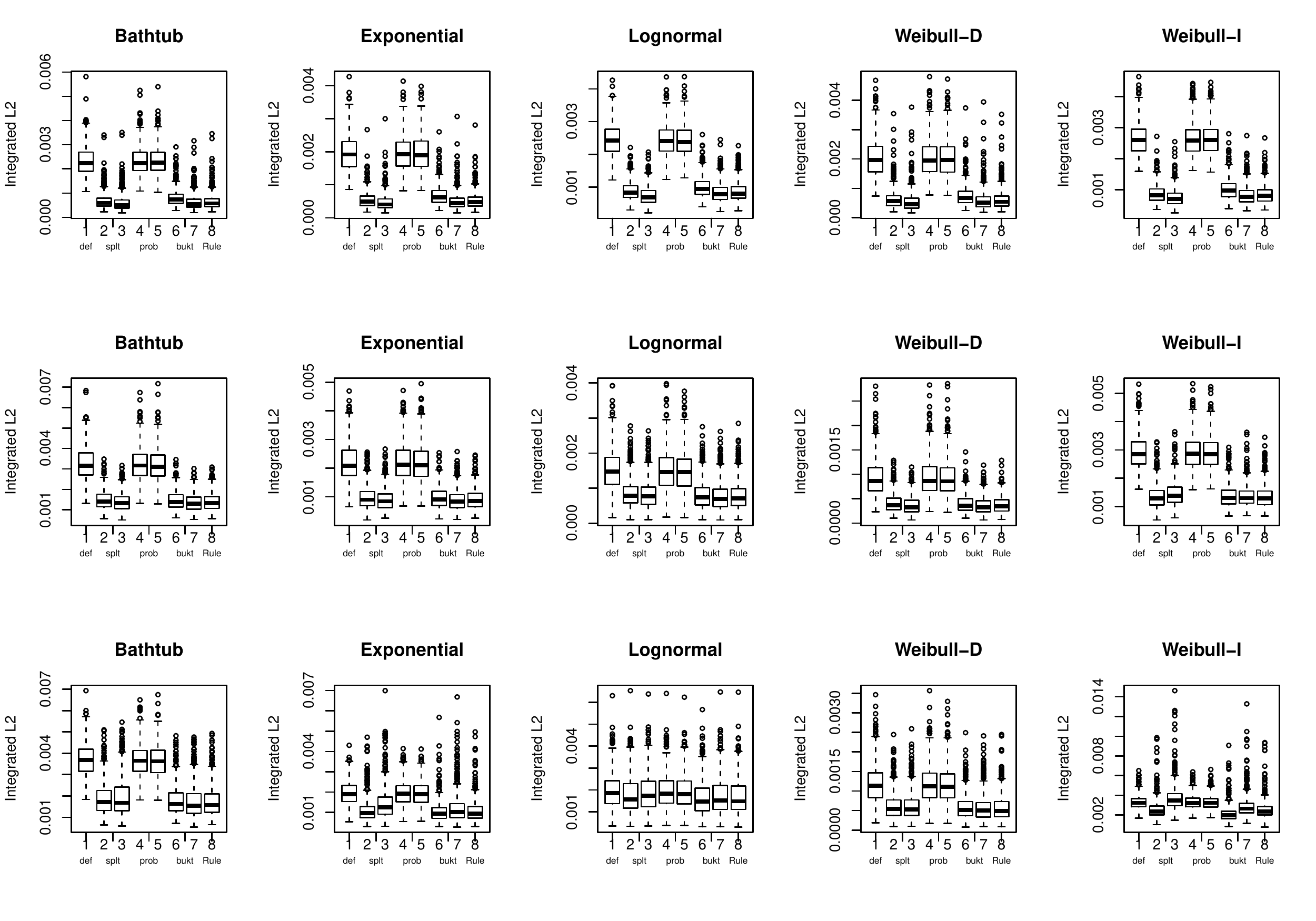}
\caption{\small{Example: integrated $L_2$ difference of IC cforest with different \textit{minsplit}, \textit{minprob} and \textit{minbucket} values, with $n = 500$, no right-censoring and the interval-censoring width generated by $G_1(t)$. 1-(\textit{minsplit}, \textit{minprob}, \textit{minbucket}) $=(20,0.01,7)$. 2-(\textit{minsplit}, \textit{minprob}, \textit{minbucket}) $=(75,0.01,7)$, 3-(\textit{minsplit}, \textit{minprob}, \textit{minbucket}) $=(100,0.01,7)$, 4-(\textit{minsplit}, \textit{minprob}, \textit{minbucket}) $=(20,0.05,7)$, 5-(\textit{minsplit}, \textit{minprob}, \textit{minbucket}) $=(20,0.10,7)$, 6-(\textit{minsplit}, \textit{minprob}, \textit{minbucket}) $=(20,0.01,30)$, 7-(\textit{minsplit}, \textit{minprob}, \textit{minbucket}) $=(20,0.01,40)$. 8-The \q{15\%-Default-6\% Rule}: (\textit{minsplit}, \textit{minprob}, \textit{minbucket}) $=(75,0.01,30)$. Top row gives results for the first setup (tree structure), middle row gives results for the second setup (linear model), and bottom row gives results for the third setup (nonlinear model).}} 
\end{figure}
\

\

\newpage

\

\

\section{Evaluation of tuning parameters for $N = 1000$} 
\subsection{\textit{mtry} as a tuning parameter}\label{sec:etN11}
\begin{figure}[H]
\includegraphics[scale=0.67]{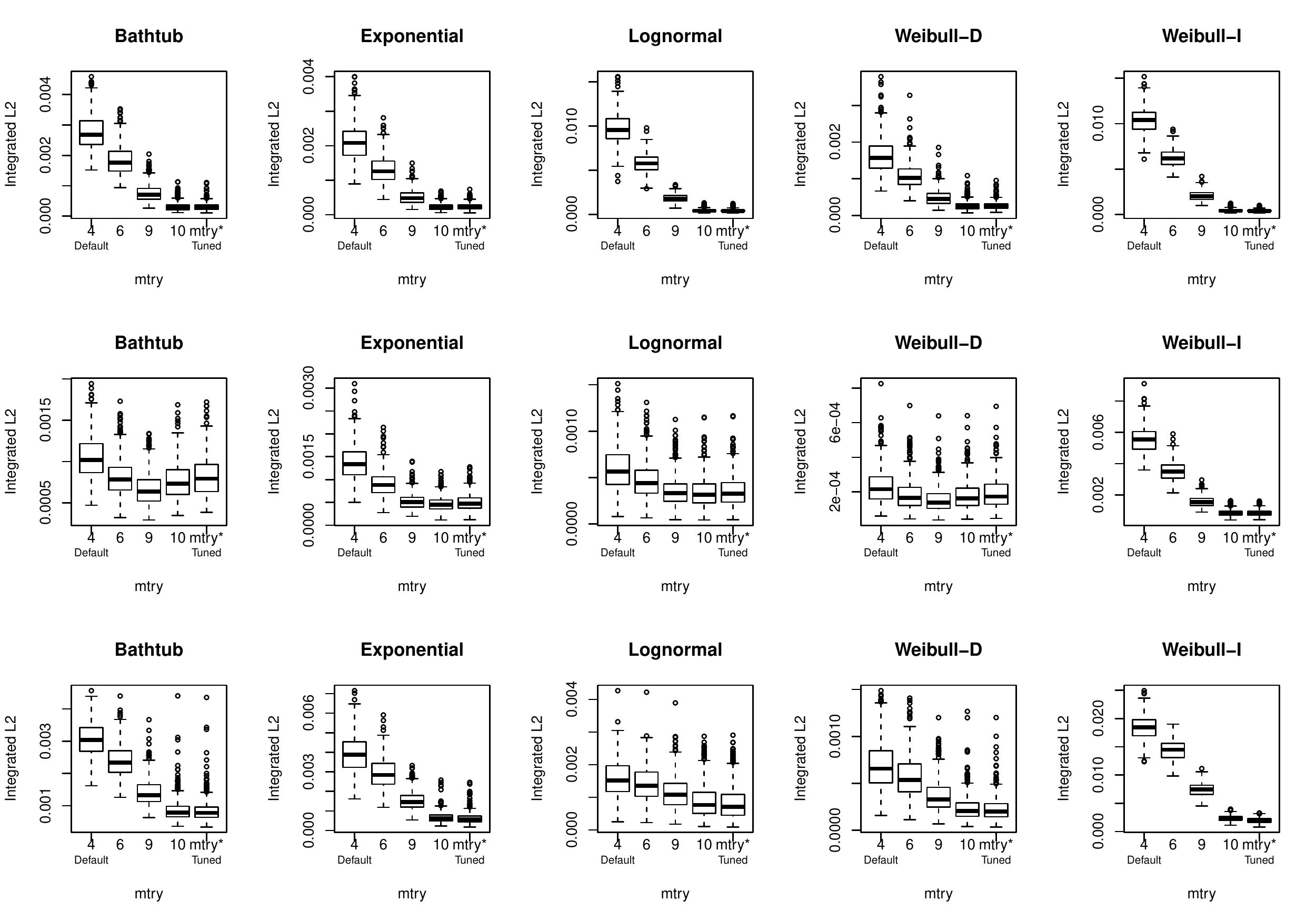}
\caption{\small{Integrated $L_2$ difference of IC cforest with different \textit{mtry} values, with $n = 1000$, no right censoring and the interval censoring width generated by $G_1(t)$. The default value in \texttt{cforest} function is $\sqrt{m} \approx 4$. The value of \textit{mtry} tuned by the \q{out-of-bag} tuning procedure is given in the last column in each boxplot. Top row gives results for the first setup (tree structure), middle row gives results for the second setup (linear model), and bottom row gives results for the third setup (nonlinear model).}}
\end{figure}

\

\

\newpage

\

\

\subsection{\textit{minsplit}, \textit{minprob} and \textit{minbucket} as tuning parameters}\label{sec:etN12}
\begin{figure}[H]
\includegraphics[scale=0.67]{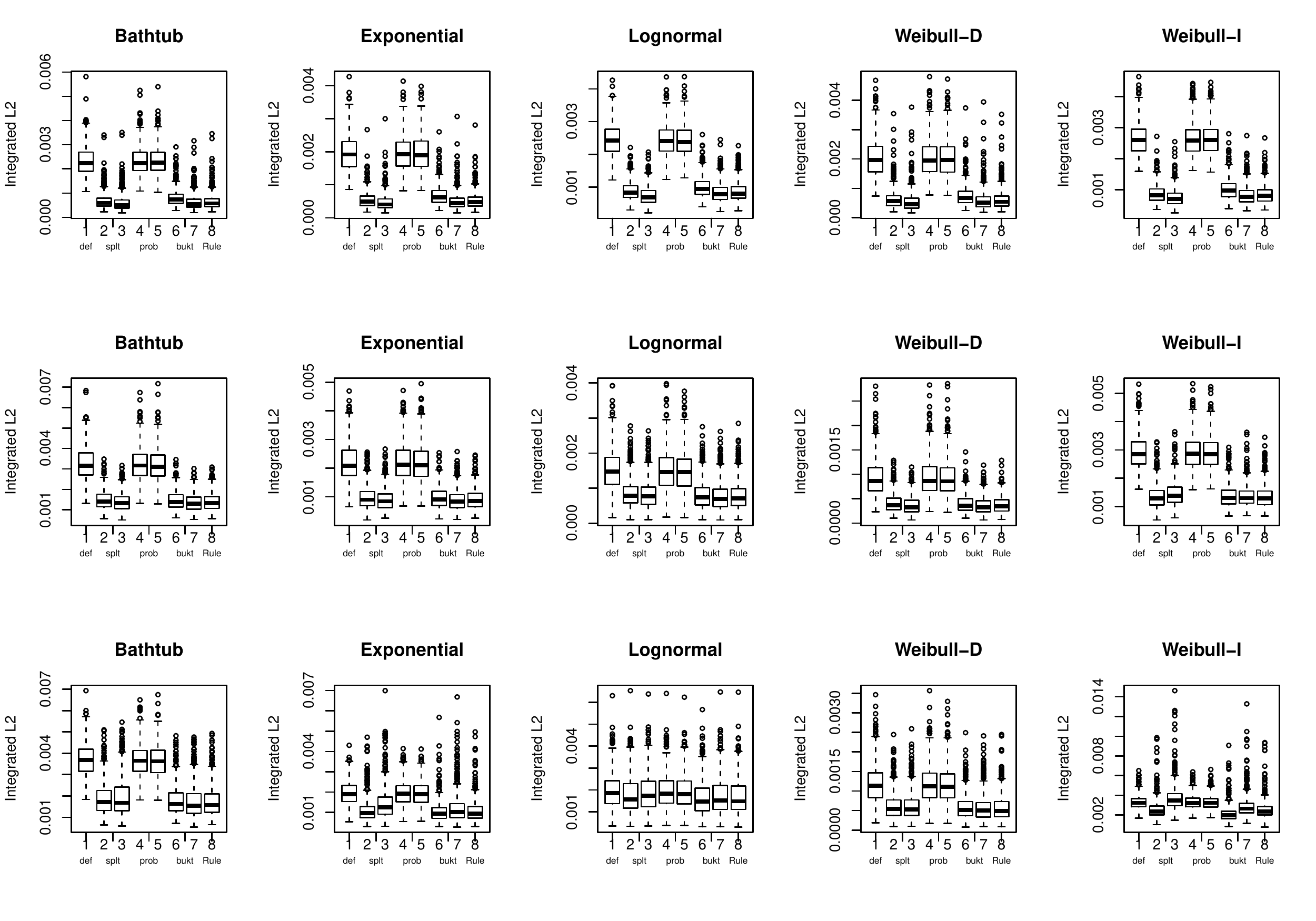}
\caption{\small{Example: integrated $L_2$ difference of IC cforest with different \textit{minsplit}, \textit{minprob} and \textit{minbucket} values, with $n = 1000$, no right-censoring and the interval-censoring width generated by $G_1(t)$. 1-(\textit{minsplit}, \textit{minprob}, \textit{minbucket}) $=(20,0.01,7)$. 2-(\textit{minsplit}, \textit{minprob}, \textit{minbucket}) $=(150,0.01,7)$, 3-(\textit{minsplit}, \textit{minprob}, \textit{minbucket}) $=(200,0.01,7)$, 4-(\textit{minsplit}, \textit{minprob}, \textit{minbucket}) $=(20,0.05,7)$, 5-(\textit{minsplit}, \textit{minprob}, \textit{minbucket}) $=(20,0.10,7)$, 6-(\textit{minsplit}, \textit{minprob}, \textit{minbucket}) $=(20,0.01,60)$, 7-(\textit{minsplit}, \textit{minprob}, \textit{minbucket}) $=(20,0.01,80)$. 8-The \q{15\%-Default-6\% Rule}: (\textit{minsplit}, \textit{minprob}, \textit{minbucket}) $=(150,0.01,60)$. Top row gives results for the first setup (tree structure), middle row gives results for the second setup (linear model), and bottom row gives results for the third setup (nonlinear model).}} 
\end{figure}
\

\

\newpage

\

\

\section{Estimation performance for $N = 500$}\label{sec:epN5}
\subsection{Method performance under three different underlying true models}
\begin{figure}[H]
\includegraphics[scale=0.67]{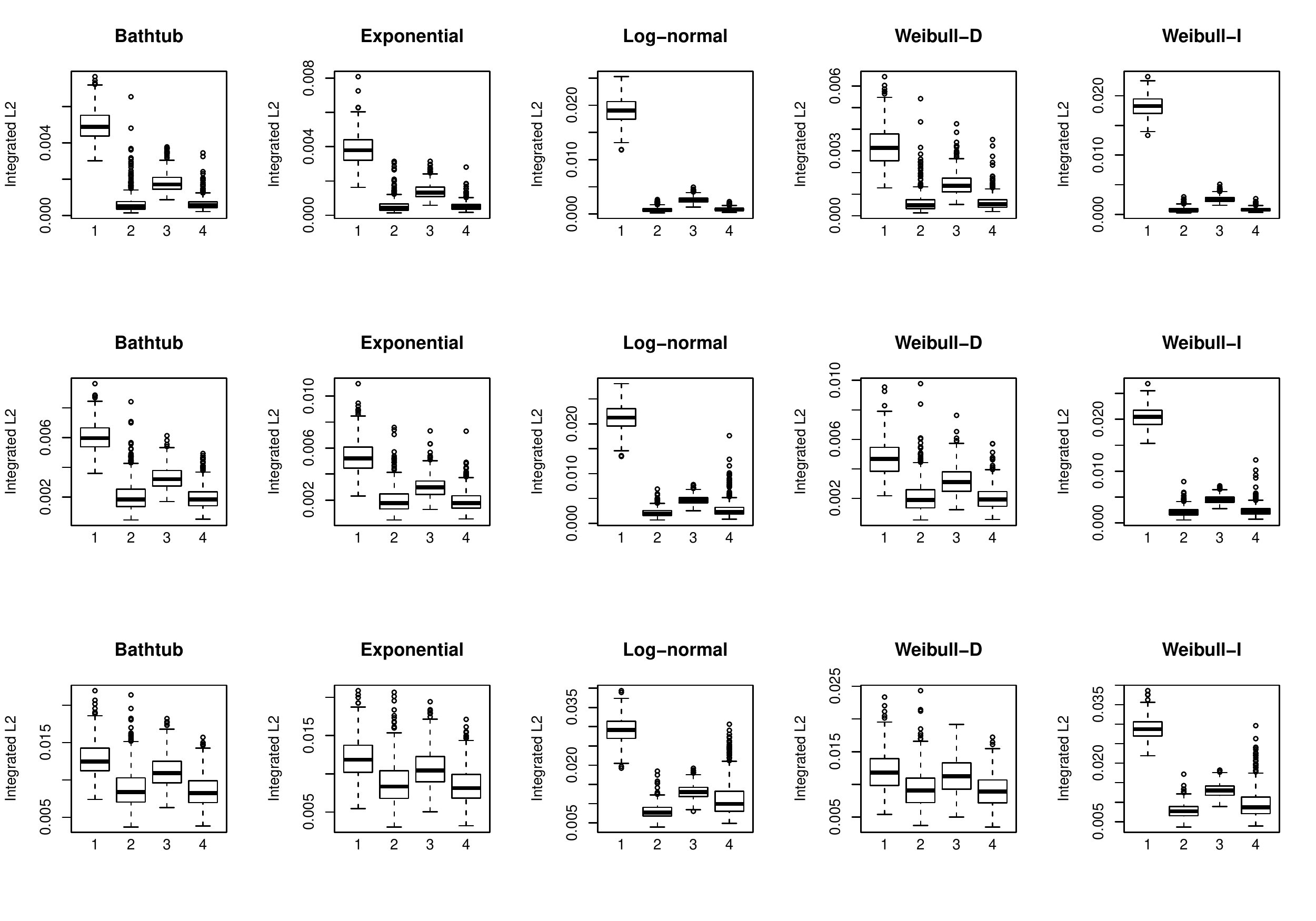}
\caption{\small{True tree model with censoring interval width generated from $G_1(t)$: integrated $L_2$ difference boxplots with $n=500$. Methods are numbered as 1-IC Cox model, 2-IC ctree, 3-IC cforest with parameters set by default, 4-IC cforest with parameters set through \q{out-of-bag} tuning procedure and the \q{15\%-Default-6\% Rule.} Top row gives results without right-censoring, middle row gives results for light (right-)censoring, and bottom row gives results for heavy (right-)censoring.}}
\end{figure}
\begin{figure}[H]
\includegraphics[scale=0.65]{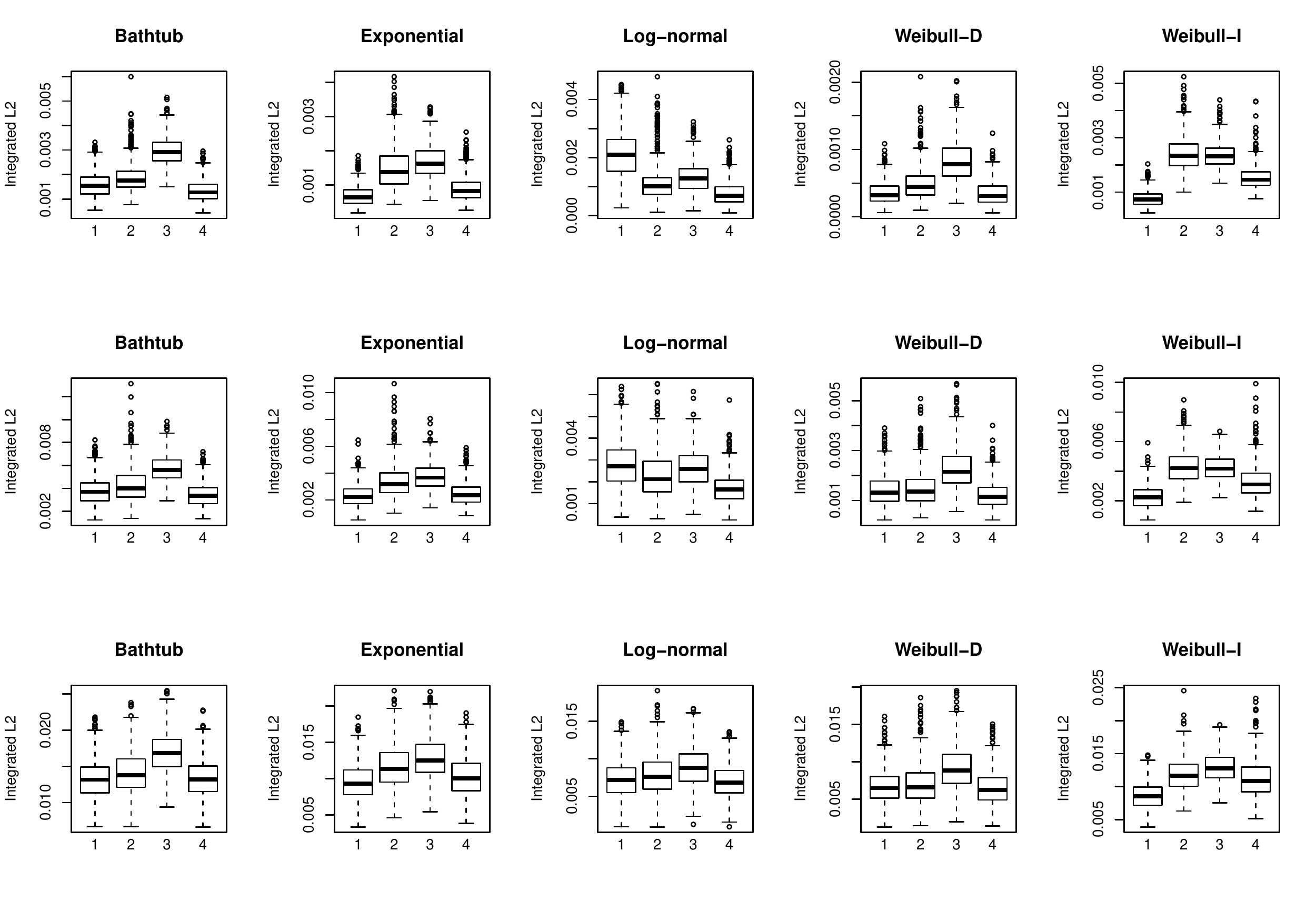}
\caption{\small{True linear model with censoring interval width generated from $G_1(t)$: integrated $L_2$ difference boxplots with $n=500$. Methods are numbered as 1-IC Cox model, 2-IC ctree, 3-IC cforest with parameters set by default, 4-IC cforest with parameters set through \q{out-of-bag} tuning procedure and the \q{15\%-Default-6\% Rule.} Top row gives results without right-censoring, middle row gives results for light (right-)censoring, and bottom row gives results for heavy (right-)censoring.}}
\end{figure}
\begin{figure}[H]
\includegraphics[scale=0.65]{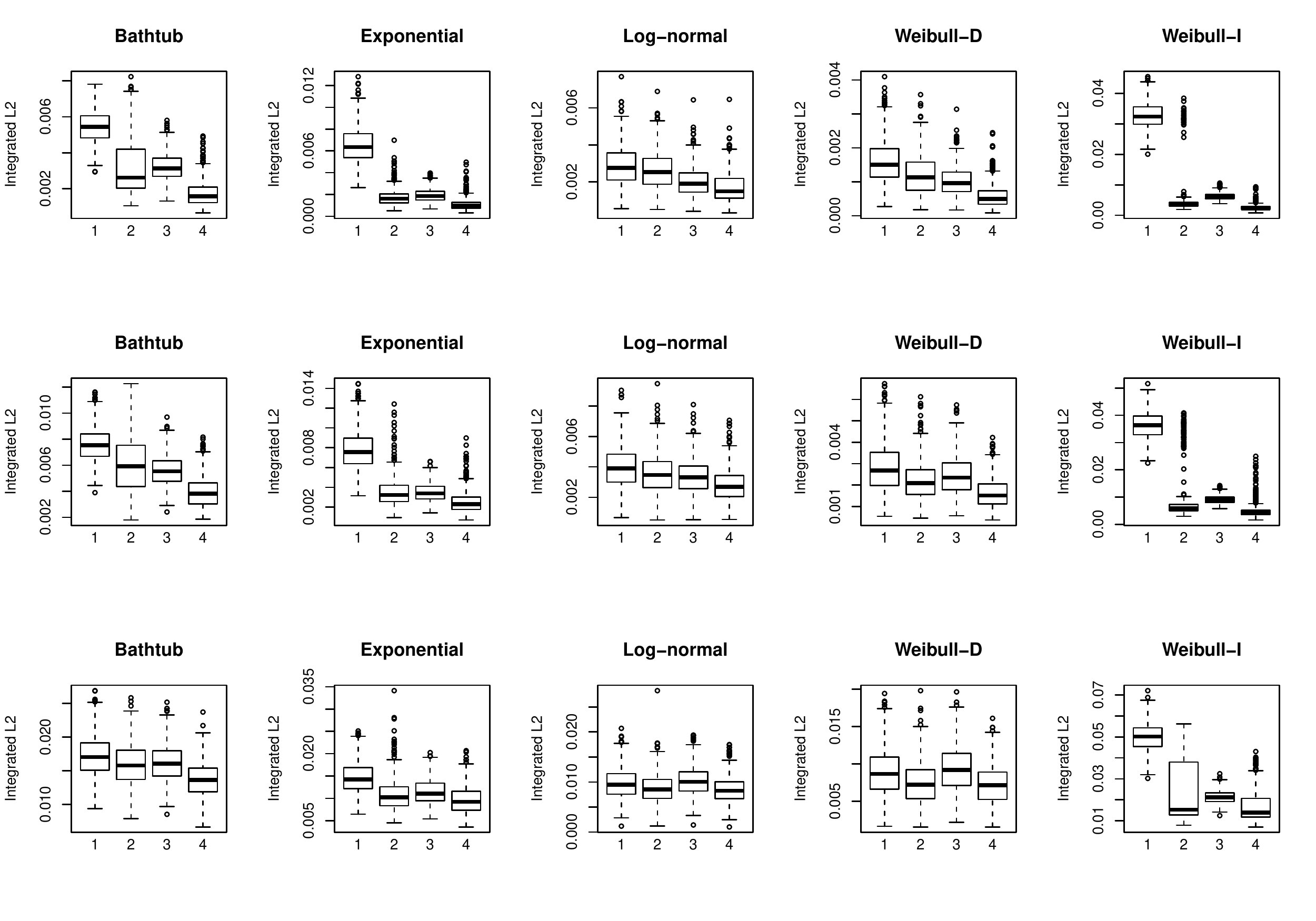}
\caption{\small{True nonlinear model with censoring interval width generated from $G_1(t)$: integrated $L_2$ difference boxplots with $n=500$. Methods are numbered as 1-IC Cox model, 2-IC ctree, 3-IC cforest with parameters set by default, 4-IC cforest with parameters set through \q{out-of-bag} tuning procedure and the \q{15\%-Default-6\% Rule.} Top row gives results without right-censoring, middle row gives results for light (right-)censoring, and bottom row gives results for heavy (right-)censoring.}}
\end{figure}
\

\

\newpage

\

\

\subsection{Method performance under different censoring interval widths}
\begin{figure}[H]
\includegraphics[scale=0.65]{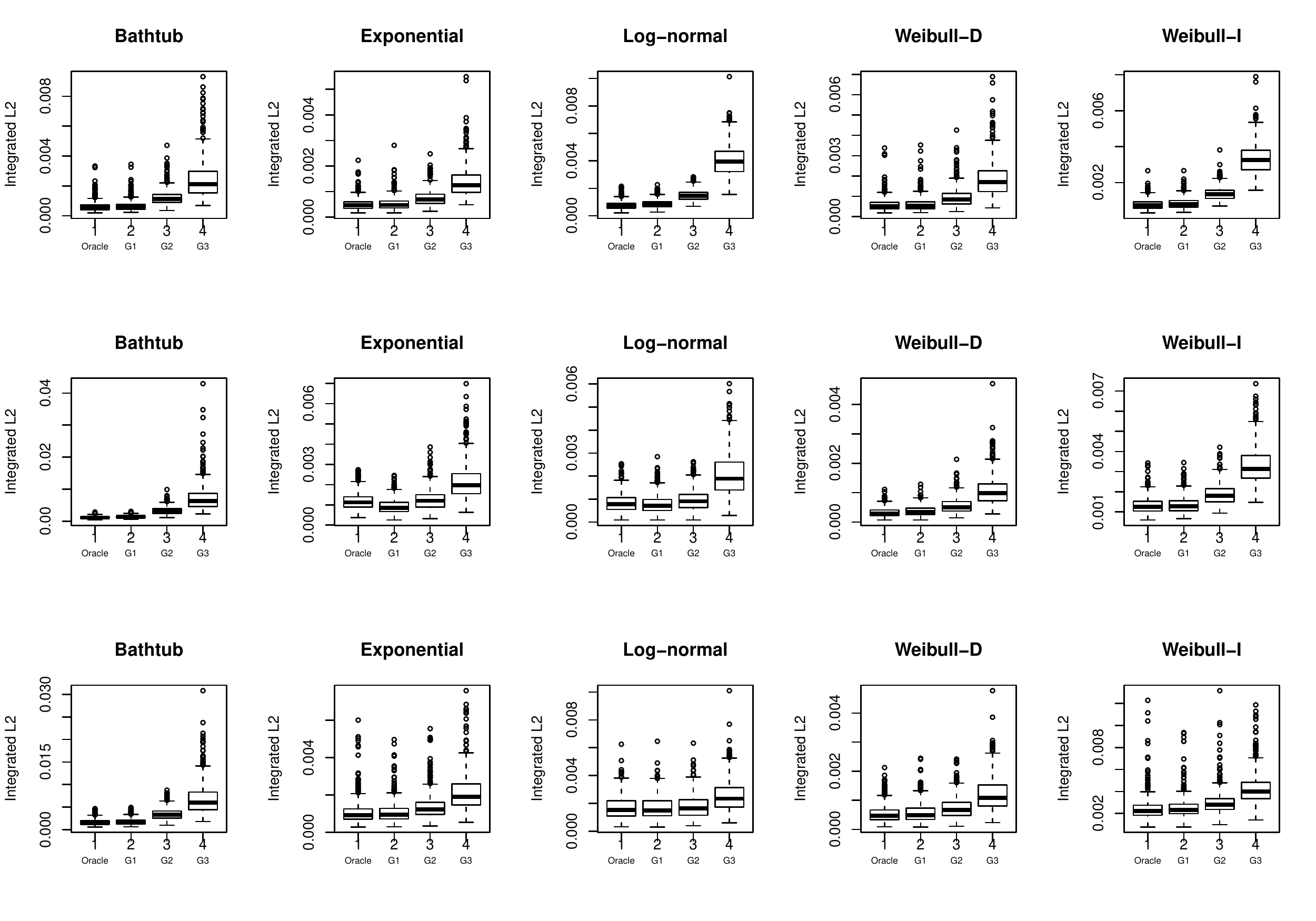}
\caption{\small{Integrated $L_2$ difference boxplots with $n=500$, no right-censoring. 1-Oracle, 2-censoring interval width generated from $G_1(t)$, 3-Censoring interval width generated from $G_2(t)$, 4-Censoring interval width generated from $G_3(t)$. Methods that give results in columns 2-4 are IC cforest with \textit{mtry} chosen through \q{out-of-bag} tuning procedure and \textit{minsplit}, \textit{minprob}, \textit{minbucket} chosen following \q{15\%-def-6\% Rule.} Top row gives results for tree model, middle row gives results for linear model, and bottom row gives results for nonlinear model.}}
\end{figure}

\begin{figure}[H]
\includegraphics[scale=0.65]{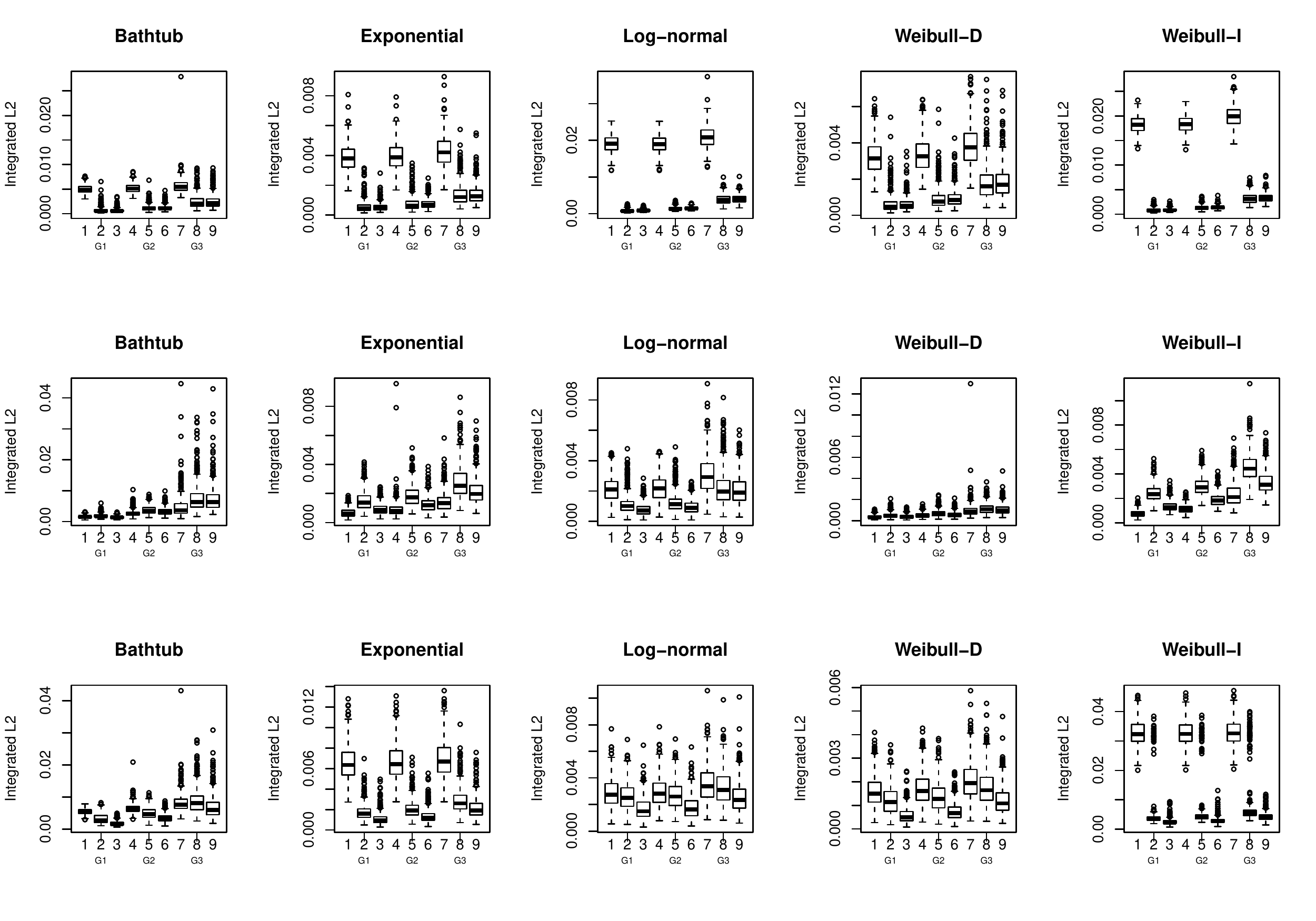}
\caption{\small{Integrated $L_2$ difference boxplots with $n=500$, no right-censoring. In each boxplot, 1-3 give results of IC Cox, IC ctree and IC cforest for censoring interval width generated from $G_1(t)$ respectively, 4-6 gives results of IC Cox, IC ctree and IC cforest for censoring interval width generated from $G_2(t)$ respectively, 7-9 give results of IC Cox, IC ctree and IC cforest for censoring interval width generated from $G_3(t)$ respectively. Top row gives results for tree model, middle row gives results for linear model, and bottom row gives results for nonlinear model.}}
\end{figure}
\

\

\newpage

\

\

\section{Estimation performance for $N = 1000$}\label{sec:epN1}
\subsection{Method performance under three different underlying true models}
\begin{figure}[H]
\includegraphics[scale=0.67]{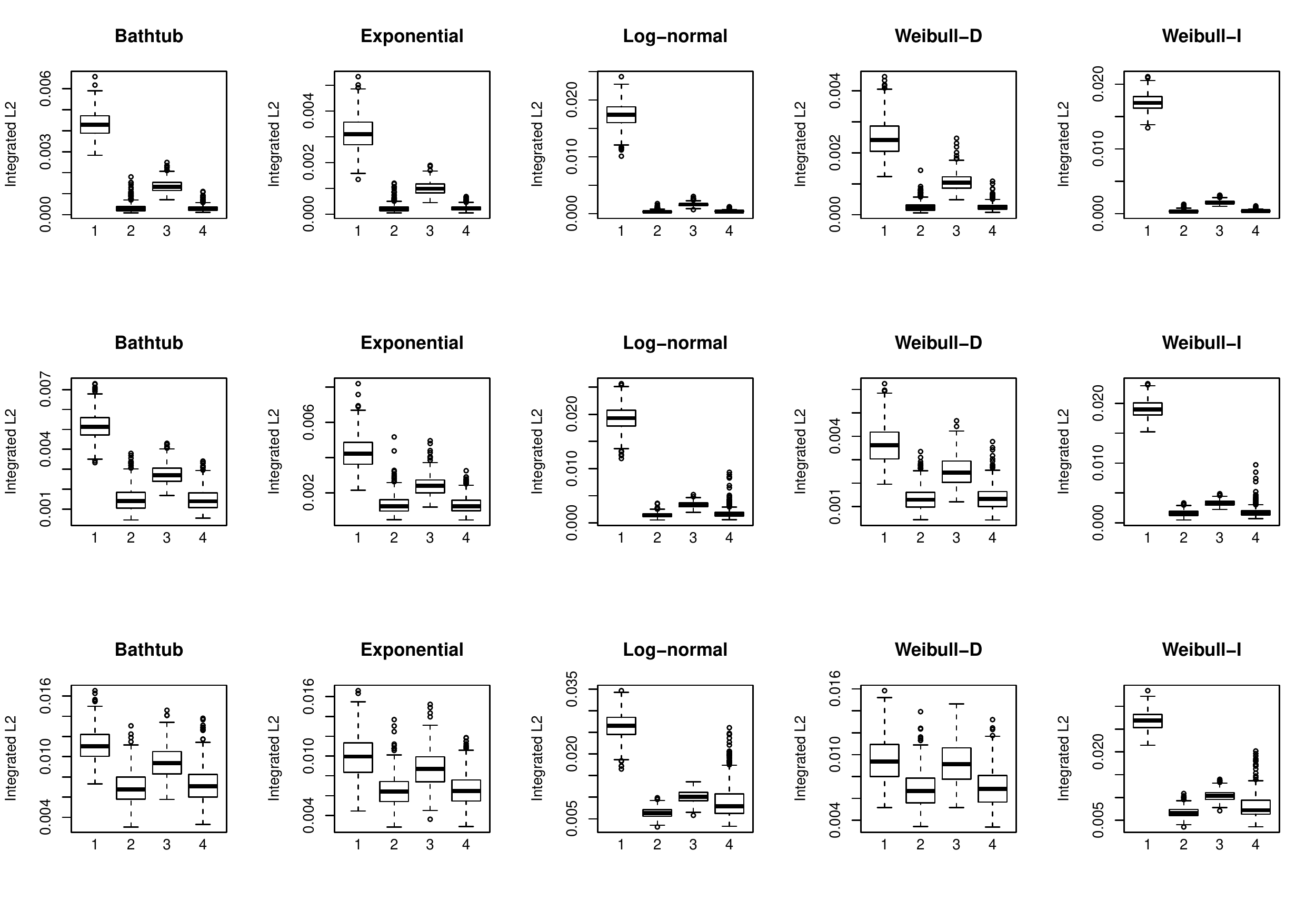}
\caption{\small{True tree model with censoring interval width generated from $G_1(t)$: integrated $L_2$ difference boxplots with $n=1000$. Methods are numbered as 1-IC Cox model, 2-IC ctree, 3-IC cforest with parameters set by default, 4-IC cforest with parameters set through \q{out-of-bag} tuning procedure and the \q{15\%-Default-6\% Rule.} Top row gives results without right-censoring, middle row gives results for light (right-)censoring, and bottom row gives results for heavy (right-)censoring.}}
\end{figure}
\begin{figure}[H]
\includegraphics[scale=0.65]{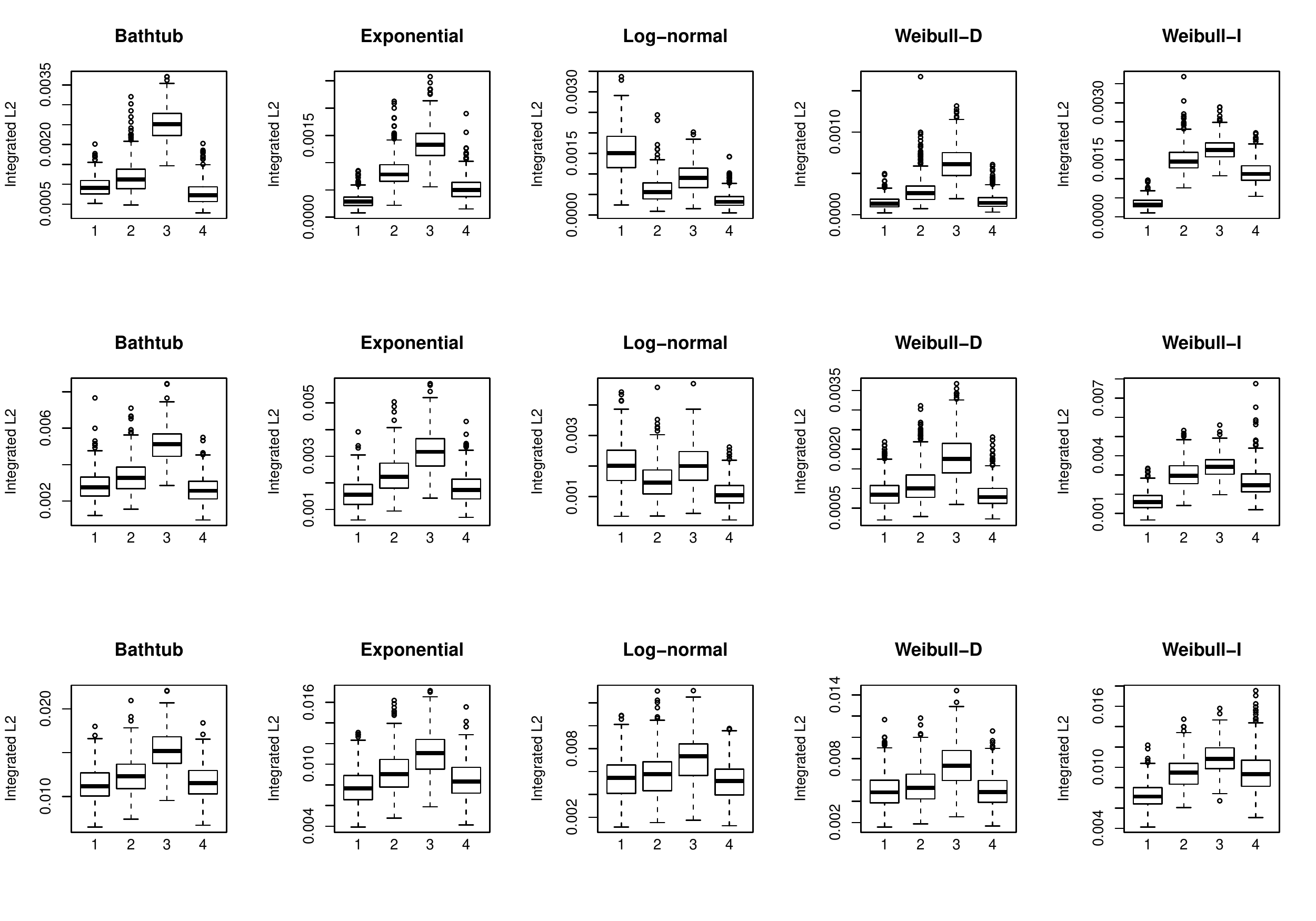}
\caption{\small{True linear model with censoring interval width generated from $G_1(t)$: integrated $L_2$ difference boxplots with $n=1000$. Methods are numbered as 1-IC Cox model, 2-IC ctree, 3-IC cforest with parameters set by default, 4-IC cforest with parameters set through \q{out-of-bag} tuning procedure and the \q{15\%-Default-6\% Rule.} Top row gives results without right-censoring, middle row gives results for light (right-)censoring, and bottom row gives results for heavy (right-)censoring.}}
\end{figure}
\begin{figure}[H]
\includegraphics[scale=0.65]{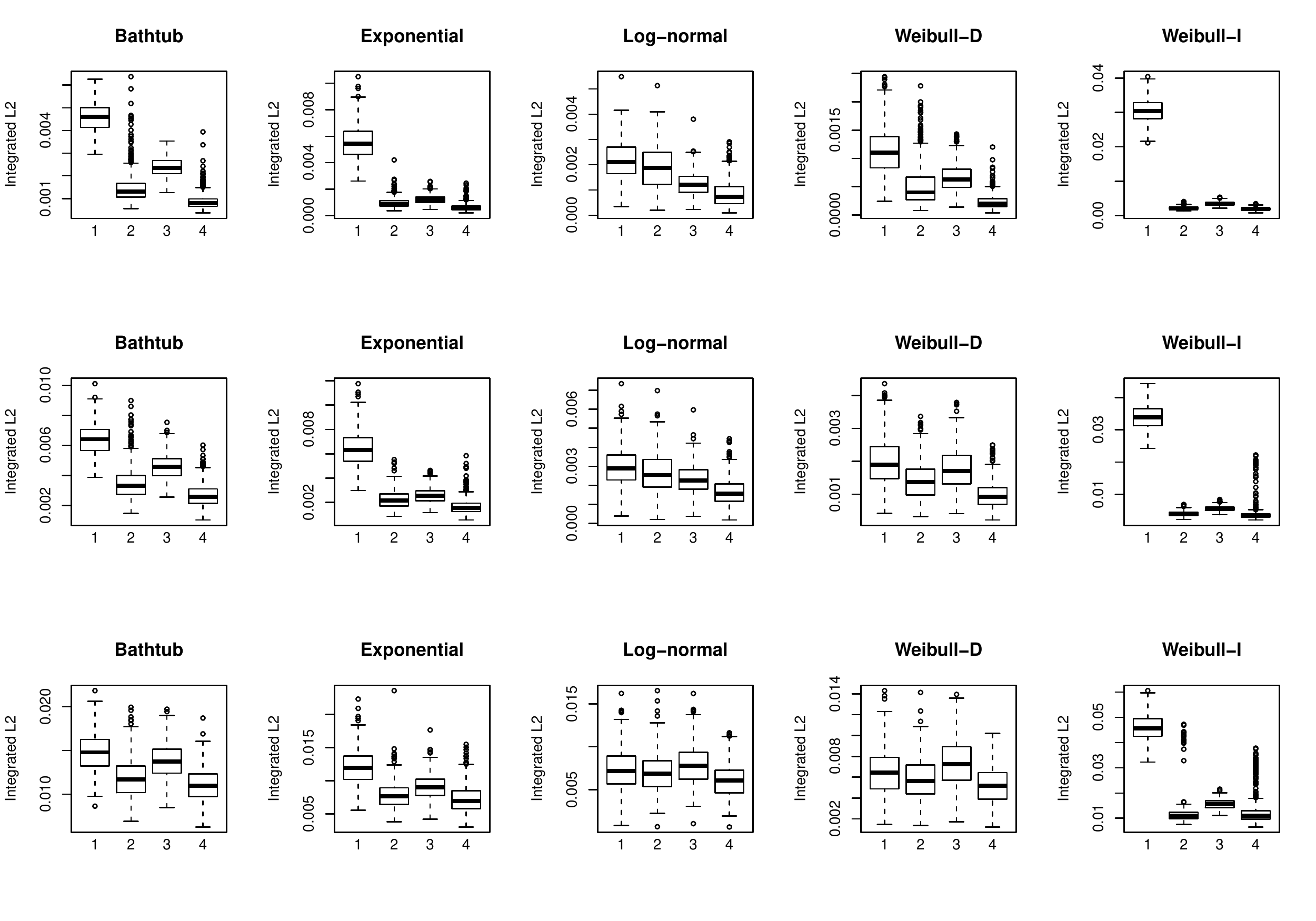}
\caption{\small{True nonlinear model with censoring interval width generated from $G_1(t)$: integrated $L_2$ difference boxplots with $n=1000$. Methods are numbered as 1-IC Cox model, 2-IC ctree, 3-IC cforest with parameters set by default, 4-IC cforest with parameters set through \q{out-of-bag} tuning procedure and the \q{15\%-Default-6\% Rule.} Top row gives results without right-censoring, middle row gives results for light (right-)censoring, and bottom row gives results for heavy (right-)censoring.}}
\end{figure}
\

\

\newpage

\

\

\subsection{Method performance under different censoring interval widths}
\begin{figure}[H]
\includegraphics[scale=0.65]{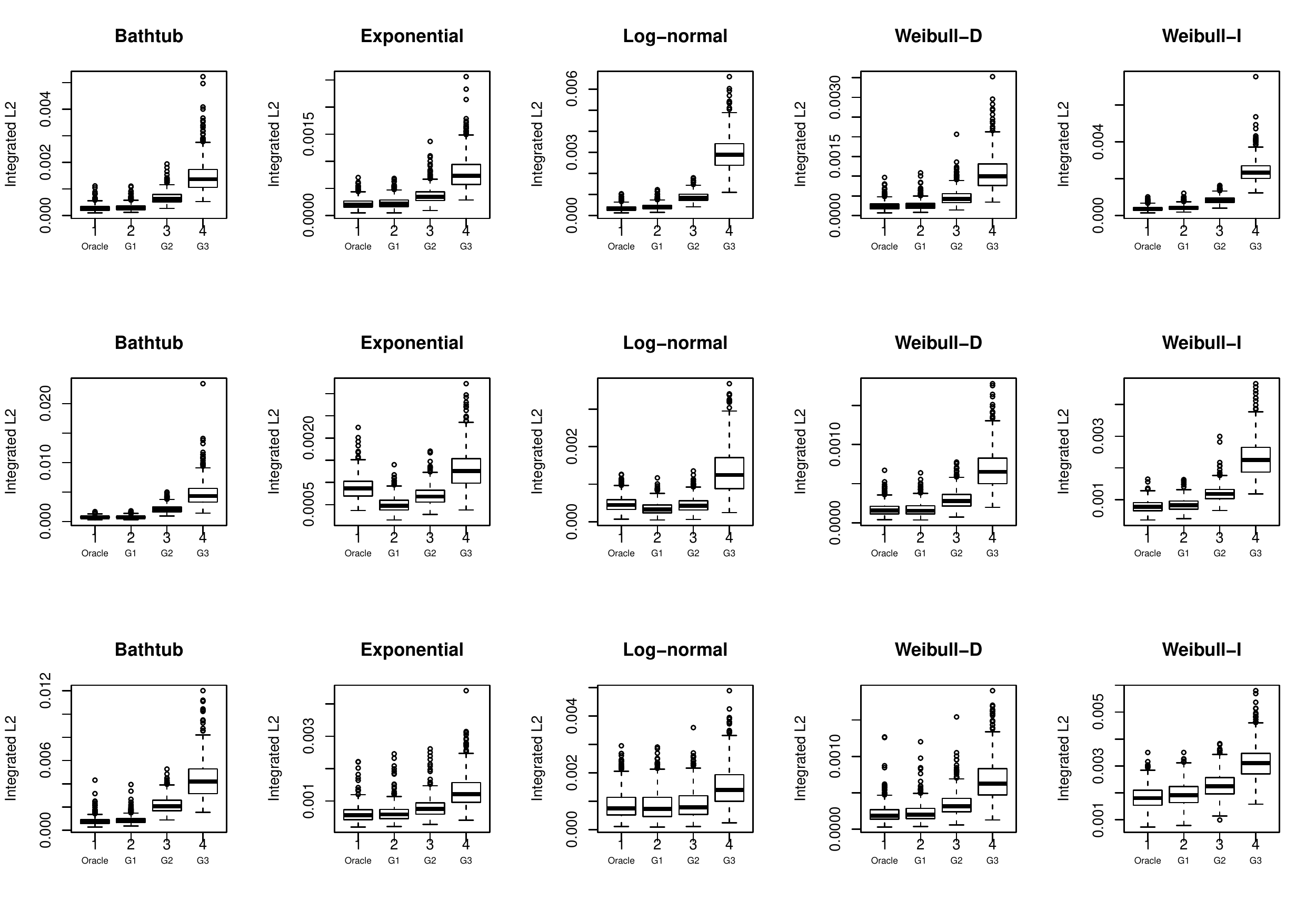}
\caption{\small{Integrated $L_2$ difference boxplots with $n=1000$, no right-censoring. 1-Oracle, 2-censoring interval width generated from $G_1(t)$, 3-Censoring interval width generated from $G_2(t)$, 4-Censoring interval width generated from $G_3(t)$. Methods that give results in columns 2-4 are IC cforest with \textit{mtry} chosen through \q{out-of-bag} tuning procedure and \textit{minsplit}, \textit{minprob}, \textit{minbucket} chosen following \q{15\%-def-6\% Rule.} Top row gives results for tree model, middle row gives results for linear model, and bottom row gives results for nonlinear model.}}
\end{figure}

\begin{figure}[t!]
\includegraphics[scale=0.65]{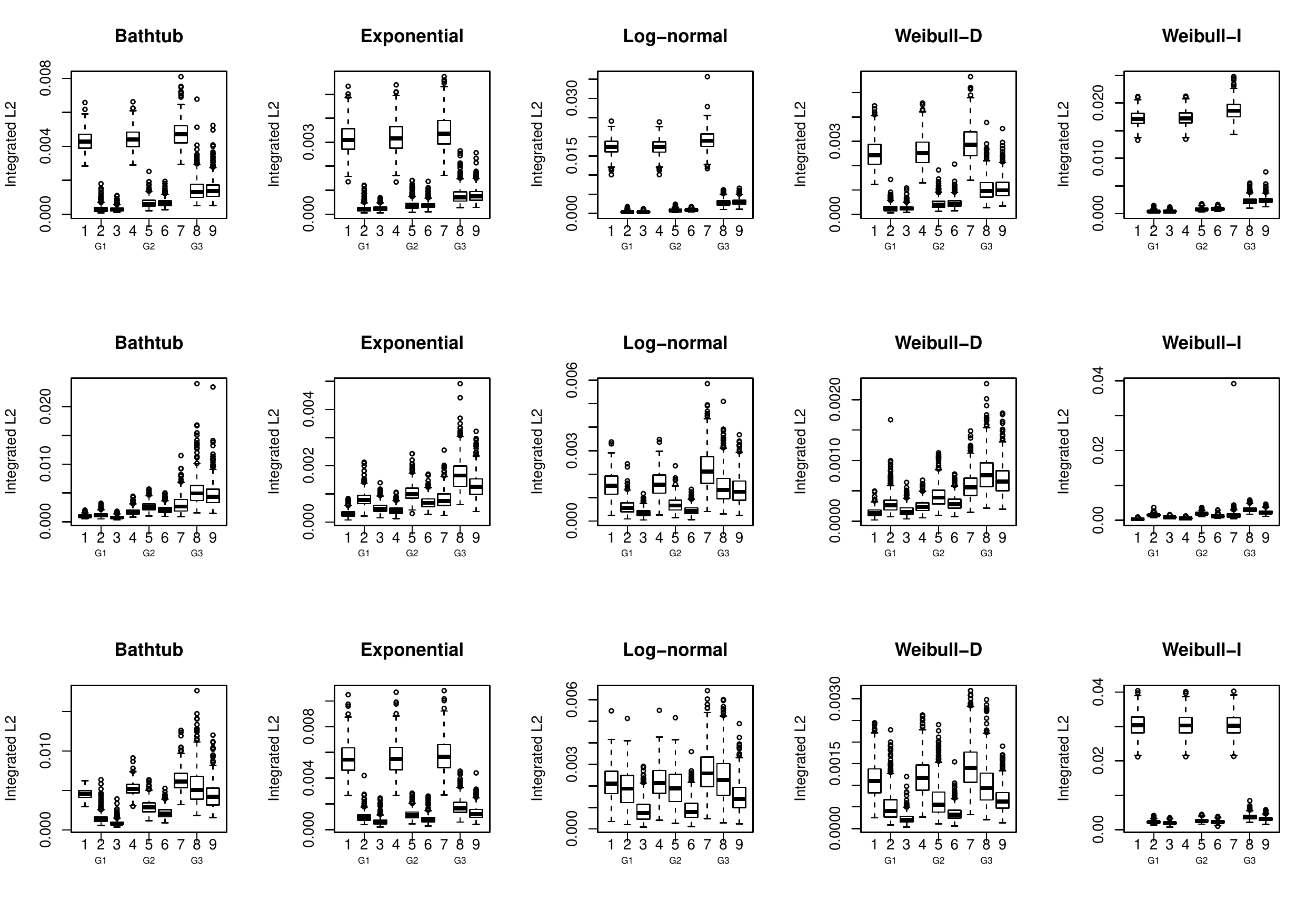}
\caption{\small{Integrated $L_2$ difference boxplots with $n=1000$, no right-censoring. In each boxplot, 1-3 give results of IC Cox, IC ctree and IC cforest for censoring interval width generated from $G_1(t)$ respectively, 4-6 gives results of IC Cox, IC ctree and IC cforest for censoring interval width generated from $G_2(t)$ respectively, 7-9 give results of IC Cox, IC ctree and IC cforest for censoring interval width generated from $G_3(t)$ respectively. Top row gives results for tree model, middle row gives results for linear model, and bottom row gives results for nonlinear model.}}
\end{figure}
\end{appendices}

\end{document}